\documentclass[twocolumn,showpacs,showkeys]{revtex4}
\usepackage{graphicx}
\input{psfig.sty}

\newcommand{\bastar}{\begin{eqnarray*}}
\newcommand{\eastar}{\end{eqnarray*}}
\newskip\humongous \humongous=0pt plus 1000pt minus 1000pt

\newif\ifdtup

\relax
\newcommand{\be}{\begin{equation}}
\newcommand{\ee}{\end{equation}}
\newcommand{\bea}{\begin{eqnarray}}
\newcommand{\eea}{\end{eqnarray}}
\newcommand{\X}{{\vec X}}
\newcommand{\pro}{\partial}
\newcommand{\n}{\hat n}
\newcommand{\oneg}{\displaystyle\frac{1}{g}}

\newcommand{\D}{{\hat D}}

\newcommand{\A}{{\vec A}}
\newcommand{\valpha}{{\vec \alpha}}

\newcommand{\dfrac}{\displaystyle\frac}
\newcommand{\ba}{\begin{array}}
\newcommand{\ea}{\end{array}}

\newcommand{\nn}{\nonumber}
\newcommand{\hn}{\hat n}
\begin{document}
\title {Non-Abelian Vortices in Condensed Matter Physics}
\author{Y. M. Cho}
\email{ymcho@yongmin.snu.ac.kr}
\affiliation{Center for Theoretical Physics, College of Natural Sciences,
Seoul National University,
Seoul 151-742, Korea  \\
and \\
C.N.Yang Institute for Theoretical Physics, State University of
New York, Stony Brook, NY 11790, USA}
\author{Hyojoong Khim and Namsik Yong}
\email{kimhj76@phya.snu.ac.kr}
\affiliation{School of Physics, College of Natural Sciences,
Seoul National University,
Seoul 151-742, Korea  \\}
\begin{abstract}
~~~~~We study the non-Abelian topological
vortices in condensed matter physics, whose topological flux 
quantum number is described by $\pi_2(S^2)$, not by $\pi_1(S^1)$.
We present two examples, a magnetic vortex in two-gap superconductor
and a vorticity vortex in two-component Bose-Einstein
condensate. In both cases the condensates exhibit a global $SU(2)$
symmetry which allows the non-Abelian topology.
We establish the non-Abelian flux quantization in two-gap superconductor
by demonstrating the existence of non-Abelian magnetic vortex 
whose flux is quantized in the unit $4\pi/g$, 
not $2\pi/g$. We also discuss a genuine non-Abelian gauge
theory of superconductivity
which has a local $SU(2)$ gauge symmetry, and establish
the non-Abelian Meissner effect in the non-Abelian superconductor.
We compare the non-Abelian
vortices with the well-known Abelian Abrikosov
vortex, and discuss how these non-Abelian vortices could be observed
experimentally in two-gap superconductor made of ${\rm MgB_2}$
and spin-1/2 condensate of $^{87}{\rm Rb}$ atoms.
Finally, we argue that the existence of
the non-Abelian vortices provides a strong
evidence for the existence of topological
knots in these condensed matters whose topology is fixed by $\pi_3(S^2)$,
which one can construct by twisting and connecting
the periodic ends of the non-Abelian vortices.
\end{abstract}
\maketitle

\section{Introduction}

Topological objects, in particular finite energy topological
solitons, have played an important role
in physics. In condensed matter the Abrikosov vortex in
ordinary superconductors and similar ones in Bose-Einstein
condensates (as well as in superfluids)
are the well-known examples of the (1+2)-dimensional
topological solitons, which have been studied extensively
theoretically and experimentally \cite{abri}.
These vortices originates from the fact that the underlying
theory has a $U(1)$ (i.e., an Abelian) symmetry, which provides
a $\pi_1(S^1)$ topology to the vortices.
On the other hand, recent experimental advances have enabled
us to create far more complex condensed matters,
such as two-gap superconductors and two-component Bose-Einstein
condensates (BEC) \cite{sc,bec}.
This opens a new possibility for us to 
observe far more interesting topological objects in 
condensed matters \cite{cho1,cho2,ruo,bat}.

This is because the new condensates, due to their multi-component 
structure, can have a complex topological structure 
which is absent in the single-component condensates.
{\it The purpose of this paper is to demonstrate the existence of
non-Abelian vortices and topological knots in 
the new condensed matters.} In the following
we present two examples of novel non-Abelian vortex
whose topology is described by $\pi_2(S^2)$,
one in two-gap superconductor and one in two-component BEC, 
which are completely different from
the well-known Abelian Abrikosov vortex. Furthermore we predict 
the existence of the totally new type of
solitons, the topological knots, in these condensates.
We argue that one can construct such a knot by
smoothly twisting the non-Abelian vortex, bending and connecting
the periodic ends of the vortex together \cite{cho1,cho2}.
The vortex ring constructed this way acquires the knot topology
$\pi_3(S^2)$, which guarantees the topological stability
of the knot. The existence of the new topological objects, of course,
is based on the fact that these new condensates
have a global $SU(2)$ symmetry which is non-Abelian.

A most important difference between the non-Abelian magnetic 
vortex and Abrikosov vortex in ordinary superconductor 
is the non-Abelian flux quantization. In two-gap 
superconductor we show the existence of magnetic vortex
whose flux is quantized in the unit $4\pi/g$, not $2\pi/g$. 
Mathematically this non-Abelian flux quantization 
has a deep reason. This follows from
the fact that the volume of the $U(1)$ 
subgroup of $SU(2)$ is twice as big as the volume of 
the Abelian $U(1)$. 

Since the superconductivity has always been described by 
an Abelian (i.e., electromagnetic) interaction
one may assume that the effective theory of two-gap superconductor
should be an Abelian gauge theory which has a global
$SU(2)$ symmetry. But in this case
the two condensates must carry the same charge, because
there is no way that the Abelian gauge field can
couple to a doublet condensate whose components
have opposite charges. But this does not mean that 
a doublet condensate carrying opposite
charges can not exhibit a superconductivity. 
In this paper we discuss an $SU(2)$ gauge
gauge theory of two-gap superconductor made of 
an oppositely charged doublet condensate, which can exhibit
a genuine non-Abelian superconductivity.
We establish a non-Abelian Meissner effect in this theory 
showing that the theory allows a non-Abelian 
magnetic vortex. As far as we understand, 
this type of non-Abelian superconductivity
has never been discussed before.

A similar non-Abelian vortex can also exist in two-component BEC.
This is because two-component BEC also has a non-Abelian structure 
similar to two-gap superconductor \cite{cho1}.
In this paper we consider two competing theories of two-component BEC,
the Gross-Pitaevskii theory and the gauge
theory of two-component BEC which has a vorticity interaction,
and show that both theories allow non-Abelian vorticity 
vortices very similar to each other.
We also show that the gauge theory of two-component BEC,
with the vorticity interaction, is closely related to 
two-gap superconductor. The only difference is that 
in two-component BEC the gauge interaction is induced one,
while in two-gap superconductor it is independent.

Finally we discuss the prototype non-Abelian vortex,
the baby skyrmion in Skyrme theory, and 
show that the non-Abelian vortices in condensed matters
are a direct generalzation of the baby skyrmion.
In fact we will see that all underlying theories of
non-Abelian vortices discussed in this paper are
closely related to the Skyrme theory.      
This observation has an important implication,
because this strongly implies the existence of topological 
knots in two-gap superconductor and two-component BEC. 
In Skyrme theory it is well-known that 
one can construct a knot by twisting the baby skyrmion
(making it periodic in $z$-coordinate) and connecting 
the periodic ends together \cite{cho01}. Exactly the same way 
we can construct a similar knot by twisting the non-Abelian
vortex and connecting the periodic ends together in
the new condensed matters \cite{cho1,cho2}. 
The twisted vortex ring made this way  
acquires the knot topology $\pi_3(S^2)$ and thus becomes 
a knot itself.

The paper is organized as follows. In Section II we present a
non-Abelian magnetic vortex in two-gap superconductor
in which the two condensates carry the same charge,
and estabilsh the non-Abelian magnetic flux quantization.
In Section III we compare the non-Abelian vortex
with the well-known Abelian Abrikosov
vortex, and discuss the differences between the two
vortices. In Section IV we establish a non-Abelian superconductivity
presenting a genuine $SU(2)$ gauge theory which could describe
a two-gap superconductor made of a doublet carrying
opposite charges. With this we demonstrate the existence of a
magnetic vortex in this non-Abelian superconductor
which is identical to the vortex in two-gap superconductor
based on Abelian gauge symmetry.
In Section V we discuss
a non-Abelian vortex in Gross-Pitaevskii theory
of two-component BEC, and identify the vortex as a vorticity vortex.
In Section VI we discuss a gauge theory of
two-component BEC which has a vorticity interaction,
and show that the theory has a non-Abelian
vortex which is very similar to the one in Gross-Pitaevskii theory.
In Section VII we show that the above non-Abelian vortices
are a straightforward generalization of 
the prototype non-Abelian vortex, the baby skyrmion
in Skyrme theory. Based on this we discuss how one can
construct a topological knot from the non-Abelian vortices
in the new condensates.
Finally in Section VIII we discuss the physical
implications of our results.

\section{Non-Abelian Magnetic Vortex in Two-gap Superconductor}

A recent development in condensed matter physics is the discovery of
two-gap supercondector made of ${\rm MgB_2}$ \cite{sc}.
The purpose of this section is to demonstrate the existence of
a non-Abelian magnetic vortex in two-gap
superconductors. The reason for this is that two-gap supercondector
is made of a doublet which form an $SU(2)$ multiplet.
In general, this type of topological vortex is possible
when one has more than one condensate in the superconductor
which has a symmetry group $G$ with
a nontrivial $\pi_2(G/H)$ where $H$ is the Abelian subgroup,
but in this paper we will concentrate on $SU(2)$
and two-gap superconductors.

Our non-Abelian vortex could be either relativistic
or non-relativistic, and appear with both Abelian and non-Abelian
gauge interaction. Let the complex doublet scalar field 
$\phi=(\phi_1,\phi_2)$ be the order parameter of a two-gap
superconductor. In the mean field approximation 
the Landau-Ginzburg free energy of the two-gap
superconductor could be expressed by \cite{maz}
\bea
&{\cal H} = \dfrac{\hbar^2}{2m_1} |(\mbox{\boldmath $\nabla$}
+ ig \mbox{\boldmath $A$}) \phi_1|^2 
+\dfrac{\hbar^2}{2m_2} |(\mbox{\boldmath $\nabla$}
+ ig \mbox{\boldmath $A$}) \phi_2|^2 \nn\\
&+ \tilde V(\phi_1,\phi_2) + \dfrac{1}{2} (\mbox{\boldmath $\nabla$} 
\times \mbox{\boldmath $A$})^2, \nn\\
&\tilde V(\phi_1,\phi_2)=-\tilde \alpha_1 \phi_1^{\dagger}\phi_1
-\tilde \alpha_2 \phi_2^{\dagger}\phi_2-\tilde \alpha_3(\phi_1^{\dagger}\phi_2
+\phi_2^{\dagger}\phi_1) \nn\\
&+ \dfrac{\tilde \beta_1}{2} (\phi_1^{\dagger} \phi_1)^2
+\dfrac{\tilde \beta_2}{2} (\phi_2^{\dagger} \phi_2)^2
+\tilde \beta_3 (\phi_1^{\dagger} \phi_1)(\phi_2^{\dagger} \phi_2),
\label{scfe1}
\eea
where $\tilde \alpha_i$ and $\tilde \beta_i$ are coupling constants.
This is an obvious generalization of the Landau-Ginzburg
free energy of ordinary superconductor, but one can simplify 
the above Hamiltonian with a proper normalization 
of $\phi_1$ and $\phi_2$ to $(\sqrt{2m_1}/\hbar) \phi_1$ and 
$(\sqrt{2m_2}/\hbar) \phi_2$, 
\bea
&{\cal H} = |(\mbox{\boldmath $\nabla$}
+ ig \mbox{\boldmath $A$}) \phi|^2 + V(\phi_1,\phi_2) \nn\\
&+ \dfrac{1}{2} (\mbox{\boldmath $\nabla$} 
\times \mbox{\boldmath $A$})^2,
\label{scfe2}
\eea
where $V$ is the normalized potential. Furthermore one may assume 
that the normalized potential has the simplest form, 
assuming that the normalized coupling constants 
satisfy $\alpha_1 \simeq \alpha_2$, $\alpha_3 \simeq 0$, 
and $\beta_1 \simeq \beta_2 \simeq \beta_3$ for simplicity.
In this case the potential reduces to  
\bea
&V=-\alpha \phi^{\dagger}\phi
+ \dfrac{\beta}{2} (\phi^{\dagger} \phi)^2.
\label{scpot}
\eea
With this simplification the Hamiltonian acquires a global
$SU(2)$ symmetry (as well as the $U(1)$ gauge symmetry).
In general the $SU(2)$ symmetry will be broken in real
two-gap superconductors, but one may still regard the $SU(2)$ symmetry 
as an approximate symmetry. In this sense it is worth
studying the $SU(2)$ symmetric potential first as 
a starting point. We will come back to a more general 
potential later.

With the above Hamiltonian one may try to obtain a non-Abelian vortex
minimizing the free energy. On the other hand, to study a static
vortex solution, one might as well start from
the following relativistic Lagrangian
\bea
&{\cal L} = - |D_\mu \phi|^2 + \mu^2 \phi^{\dagger}\phi
- \dfrac{\lambda}{2} (\phi^{\dagger} \phi)^2
- \dfrac{1}{4} F_{\mu \nu}^2, \nn\\
&D_\mu \phi = (\partial_\mu + ig A_\mu) \phi.
\label{sclag}
\eea
In the static limit the Lagrangian reproduces the above Hamiltonian,
and gives us an identical equation of motion.
So it must be clear that one can either start from the 
non-relativistic Hamiltonian (\ref{scfe2}) or 
the relativistic Lagrangian (\ref{sclag}) to discuss
the static solution. With this observation
we will use the Lagrangian (\ref{sclag})
to describe the two-gap superconductor in the following.
The advantage of the Lagrangian approach,
of course, is that it establishes the non-Abelian vortex of two-gap
superconductor as a topological
soliton of a renormalizable quantum field theory.

The Lagrangian has the equation of motion
\bea
&D^2\phi =\lambda(\phi^{\dagger} \phi
-\dfrac{\mu^2}{\lambda})\phi, \nn\\
&\partial_\mu F_{\mu \nu} = -j_\nu = ig \Big[(D_\nu
\phi)^{\dagger}\phi - \phi ^{\dagger}(D_\nu \phi) \Big].
\label{sceq1}
\eea
Now, with
\bea
&\phi =\dfrac{1}{\sqrt 2} \rho \zeta,~~~~~|\phi|=\dfrac{\rho}{\sqrt 2},
~~~~~{\zeta}^{\dagger}\zeta = 1, \nn\\
&\hat n = \zeta^{\dagger} \vec \sigma \zeta,
~~~~~\hn^2=1,
\label{def}
\eea
we have the following identity
\bea
&\Big(\partial_\mu -ig \tilde A_\mu - \dfrac{1}{2} \vec \sigma
\cdot \pro_\mu \n \Big) \zeta=0, \nn\\
&\tilde A_\mu = -\dfrac{i}{g} \zeta^{\dagger}\partial _\mu \zeta.
\label{cid}
\eea
With this we can reduce (\ref{sceq1}) to
\bea
&\partial ^2 \rho - \Big( \dfrac{1}{4}
(\partial _\mu \hat n)^2 + g^2 B_\mu^2 \Big) \rho = \dfrac{\lambda}{2}
(\rho^2-\rho_0^2)\rho, \nn\\
&\Big\{ (\partial_\mu \hat n)^2 \nn\\
&+ \Big(\partial^2 \hat n
+ \big(2\dfrac{\partial_\mu \rho}{\rho} + 2ig B_\mu \big)
\partial_\mu \hat n \Big) \cdot \vec \sigma \Big\} \zeta =0, \nn\\
&\partial_\mu F_{\mu\nu} = -j_\nu = g^2 \rho^2 B_\nu, \nn\\
&B_\mu =A_\mu + \tilde A_\mu.
\label{sceq2}
\eea
Notice that the second equation has the form
\bea
& (A+\vec B \cdot \vec \sigma) \zeta =0, \nn\\
& A=(\partial_\mu \hat n)^2, \nn\\
&\vec B = \partial^2 \hat n
+ \big(2\dfrac{\partial_\mu \rho}{\rho} + 2ig B_\mu \big)
\partial_\mu \hat n,
\eea
which is equivalent to 
\bea
&A+ \vec B \cdot \hn=0, \nn\\
& \hn \times \vec B - i \hn \times (\hn \times \vec B) =0.
\eea
So the second equation of (\ref{sceq2}) 
can be transformed to an equation for $\hn$. 
With this we can express (\ref{sceq2}) as
\bea
&\partial ^2 \rho - \Big(
\dfrac{1}{4} (\partial _\mu \hat n)^2+ g^2
B_\mu ^2 \Big) ~\rho
= \dfrac{\lambda}{2} (\rho^2-\rho_0^2)~\rho, \nn\\
&\hat n \times \partial ^2 \hat n + 2 \dfrac{\partial_\mu \rho}{
\rho} \hat n \times \partial_\mu \hat n - \dfrac{2}{g\rho^2}
(\partial_\mu F_{\mu\nu}) \partial_\nu \hat n =0, \nn\\
&\partial_\mu F_{\mu\nu} = g^2 \rho^2 B_\nu.
\label{sceq3}
\eea
This is the equation for two-gap superconductor \cite{cho2}.
The second and third equations assure that the theory
has two conserved currents, one $SU(2)$ current and
one $U(1)$ current. 

Notice that the equation (\ref{sceq2}) is written in terms of 
the $SU(2)$ doublet $\zeta$, whose target space is $S^3$.
But the equation (\ref{sceq3}) is written completely 
in terms of $\hn$, whose target space is $S^2$. 
This is made possible because of the Abelian gauge invariance.  
The Abelian gauge invariance reduces
the physical target space of $\zeta$ to the gauge orbit space
$S^2 = S^3/S^1$, which forms a $CP^1$ space which
is identical to the target space of $\n$.
In fact with (\ref{def}) the Lagrangian (\ref{sclag}) can
be expressed in terms of $\n$ (with $\rho$ and $B_\mu$)
\bea
&{\cal L} = - \dfrac{1}{2} (\pro_\mu \rho)^2 
+ \dfrac{1}{2} \rho^2 \big(\dfrac{1}{4} (\partial_\mu \hat n)^2 
+ g^2 B_\mu^2 \big) \nn\\
&+ \dfrac{1}{2} \mu^2 \rho^2
- \dfrac{\lambda}{8} \rho^4
- \dfrac{1}{4} (G_{\mu \nu}- \tilde F_{\mu \nu})^2, \nn\\
&G_{\mu\nu}= \partial _\mu B_\nu -\partial_\nu B_\mu
=F_{\mu \nu}+ \tilde F_{\mu \nu}, \nn\\
&\tilde F_{\mu \nu}= \dfrac{1}{2g} \hn \cdot (\partial_\mu \hn
\times \partial_\nu \hn). 
\label{sclag1}
\eea
Notice that the Lagrangian reproduces (\ref{sceq3}).

To obtain the vortex solution let $(\varrho,\varphi,z)$ be 
the cylindrical coordinates and choose the following ansatz
\bea
&\rho=\rho(\varrho), \nn\\
&\zeta = \Bigg( \matrix{\cos \dfrac{f(\varrho)}{2} \exp (-im\varphi) \cr
\sin \dfrac{f(\varrho)}{2} } \Bigg), \nn\\
&A_\mu= \dfrac{m}{g} A(\varrho) \partial_\mu \varphi.
\label{scans}
\eea
With this we have
\bea
&\n= \zeta^\dag \vec \sigma \zeta
= \Bigg(\matrix{\sin{f(\varrho)}\cos{m\varphi} \cr
\sin{f(\varrho)}\sin{m\varphi} \cr \cos{f(\varrho)}}\Bigg), \nn\\
&\tilde A_\mu = -\dfrac{m}{2g} \big(\cos{f(\varrho)}+1 \big)
\partial_\mu \varphi,
\eea
so that (\ref{sceq3}) is reduced to
\bea
&\ddot{\rho}+\dfrac{1}{\varrho}\dot\rho
- \Big[\dfrac{1}{4}\Big(\dot{f}^2+\dfrac{m^2}{\varrho^2}\sin^2{f}\Big) \nn\\
&+ \dfrac{m^2}{\varrho^2} \Big(A-\dfrac{\cos{f}+1}{2}\Big)^2\Big]\rho
= \dfrac{\lambda}{2}(\rho^2-\rho_0^2)\rho, \nn\\
&\ddot{f} + \Big(\dfrac{1}{\varrho}+2\dfrac{\dot{\rho}}{\rho} \Big)\dot{f}
- 2\dfrac{m^2}{\varrho^2} \Big(A-\dfrac{1}{2} \Big) \sin{f} =0, \nn\\
&\ddot{A}-\dfrac{\dot{A}}{\varrho} -g^2 \rho^2
\Big(A-\dfrac{\cos{f}+1}{2}\Big) = 0.
\label{sceq4}
\eea
In terms of
\bea
B_\mu=\dfrac{m}{g} B ~\partial_\mu\varphi
= \dfrac{m}{g} \Big(A-\dfrac{\cos{f}+1}{2} \Big)~\partial_\mu\varphi, \nn
\eea
this can be written as
\bea
&\ddot{\rho}+\dfrac{1}{\varrho}\dot\rho
- \Big[\dfrac{1}{4}\Big(\dot{f}^2+\dfrac{m^2}{\varrho^2}\sin^2{f}\Big)+
\dfrac{m^2}{\varrho^2} B^2\Big]\rho \nn\\
&=\dfrac{\lambda}{2}(\rho^2-\rho_0^2)\rho, \nn\\
&\Big(1+\dfrac{m^2}{g^2 \varrho^2} 
\dfrac{\sin^2{f}}{\rho^2} \Big) \ddot{f} \nn\\
&+ \Big(\dfrac{1}{\varrho}+ 2\dfrac{\dot{\rho}}{\rho}
+\dfrac{m^2}{g^2 \varrho^2} \dfrac{\sin{f}\cos{f}}{\rho^2} \dot{f}
- \dfrac{m^2}{g^2 \varrho^3}\dfrac{\sin^2{f}}{\rho^2} \Big) \dot{f} \nn\\
&- \dfrac{m^2}{\varrho^2}\sin{f}\cos{f} 
= \dfrac{2m^2}{g^2 \varrho^2}\Big(\ddot{B}
-\dfrac{\dot{B}}{\varrho}\Big) \dfrac{\sin{f}}{\rho^2}, \nn\\
&\ddot{B}-\dfrac{\dot{B}}{\varrho} -g^2 \rho^2 B
= \dfrac{1}{2} \Big(\ddot{f} + \dfrac{\cos{f}}{\sin{f}} \dot{f}^2 \nn\\ 
&-\dfrac{1}{\varrho} \dot{f} \Big) \sin{f}.
\label{sceq5}
\eea
Now, we impose the following boundary condition for the non-Abelian
vortex,
\bea
&\rho (0) = 0,~~~\rho(\infty) = \rho_0,
~~~f (0) = \pi,~~~f (\infty) = 0, \nn\\
& A (0) = -1,~~~A (\infty) = 1.
\label{scbc}
\eea
This need some explanation, because the boundary
value $A(0)$ is chosen to be $-1$, not $0$. This is to
assure the smoothness of the scalar field $\rho(\varrho)$
at the origin. Only with this boundary value $\rho$,
with $\rho(0)=0$, becomes analythic at the origin.
At this point one might object the boundary condition, because it creates
an apparent singularity in the gauge potential at the origin.
But notice that this singularity is an unphysical
(coordinate) singularity which can easily be removed by a gauge transformation.
In fact the singularity disappears with the gauge transformation
\bea
\phi \rightarrow \phi \exp(-im\varphi),
~~~~~A_\mu \rightarrow A_\mu + \dfrac{m}{g} \partial_\mu \varphi,
\eea
which simultaneously changes the boundary condition $A(0)=-1,~A(\infty)=1$
to $A(0)=0,~A(\infty)=2$. This boundary condition will
have an important consequence in the following.

Notice that the magnetic field $H$ of the vortex is expressed as
\bea
H= \dfrac{m}{g} \dfrac{\dot A}{\varrho}.
\eea
So we can deduce from (\ref{sceq4}) that asymptotically
the scalar field $\rho$ and the magnetic field $H$ approach
the asymptotic values $\rho_0$ and zero with the following
exponential damping
\bea
&\rho_0 - \rho \simeq \exp{(-\sqrt 2 \mu \varrho)},  \nn\\
& H \simeq \exp{(-g \rho_0 \varrho)},
\eea
respectively. This tells that, just like the single-component superconductor,
the coherence length of the condensate $\xi_C$ and
the penetration length of the magnetic field $\lambda_H$ are given by
\bea
&\xi_C = \dfrac{1}{\sqrt 2 \mu},
~~~~~\lambda_H = \dfrac{1}{\sqrt 2 \mu} \dfrac{ \sqrt \lambda}{g},  \nn\\
&\dfrac{\lambda_H}{\xi_C} = \dfrac{\sqrt \lambda}{g}.
\label{coleng}
\eea
So, when $\sqrt \lambda$ is smaller (larger) than $g$,
the superconductor becomes type I (type II).
Clearly, the condition
\bea
\sqrt {\lambda}=g,
\label{ccon}
\eea
is the critical condition
at which the superconductor changes its type from I to II.

\begin{figure}
    \includegraphics[scale=0.7]{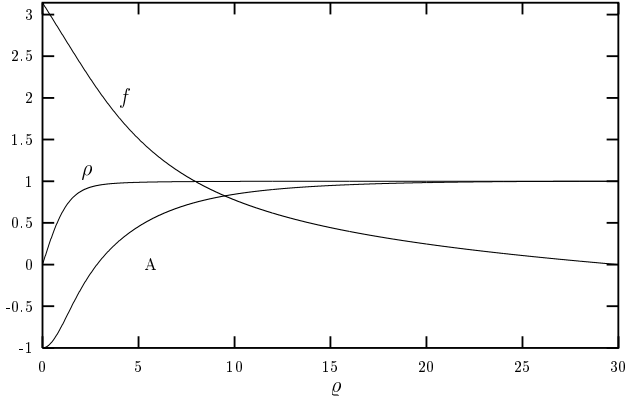}
    \caption{The non-Abelian vortex with $m=1$ in two-gap superconductor
which has $4\pi/g$ flux.
Here we have put $g=1$ and $\sqrt \lambda=2$, 
and $\varrho$ is in the unit of $\rho_0$.}
    \label{scv}
\end{figure}

With the boundary condition we can integrate (\ref{sceq4})
and obtain the non-Abelian vortex solution
of the two-gap superconductor, which is shown in Fig.\ref{scv}.
Notice that the non-trivial profile of $f(\varrho)$ 
assures that the doublet $\xi$
starts from the second component at the origin and ends up with
the first component at the infinity.
This assures that the vortex is essentially non-Abelian.

Clearly the magnetic field $H$ of the vortex
has total flux given by
\bea
&\hat \phi = \dfrac{}{} \int H d^2x
= \dfrac{2\pi m}{g} \big[A(\infty) - A(0)\big] \nn\\
&= \dfrac{4\pi m}{g}.
\label{scflux}
\eea
Notice that the unit of the non-Abelian flux is $4\pi/g$,
not $2\pi/g$. Obviously this is a direct
consequence of the boundary condition
$A(0)=-1$ (or more precisely $A(\infty)-A(0)=2$)
in (\ref{scbc}) that we discussed before.

With the ansatz (\ref{scans}) one can express the Hamiltonian
of the vortex as
\bea
&{\cal H} = \dfrac{1}{2} \int \Big[ \big(\dot \rho
\pm \dfrac{m}{\varrho}(A-\dfrac{\cos f +1}{2}) \rho \big)^2 \nn\\
&+(\dot f \pm \dfrac{m}{\varrho} \sin f)^2 \dfrac{\rho^2}{4}
+ \big(H \pm \dfrac{\sqrt \lambda}{2}(\rho^2-\rho_0^2)\big)^2 \nn\\
&\pm H \big( g\rho^2 + \sqrt \lambda (\rho_0^2-\rho^2) \big) \Big] d^2x,
\label{scham}
\eea
so that the Hamiltonian has a minimum value when
\bea
&\dot \rho \pm \dfrac{m}{\varrho}(A-\dfrac{\cos f +1}{2}) \rho =0, \nn\\
&\dot f \pm \dfrac{m}{\varrho} \sin f =0, \nn\\
&H \pm \dfrac{\sqrt \lambda}{2}(\rho^2-\rho_0^2) =0.
\label{scfoeq}
\eea
Furthermore, when the coupling constant $\lambda$ has the critical value
(\ref{ccon}), one can integrate the second order equation (\ref{sceq4}) to
the above first order equation (\ref{scfoeq}).

Integrating the second equation of (\ref{scfoeq}) we have
\bea
\cos f(\varrho) = \dfrac{\varrho^{2m} - a^2}{\varrho^{2m} + a^2},
~~~\sin f(\varrho) = \dfrac{2a\varrho^m}{\varrho^{2m} + a^2},
\label{bssol}
\eea
where $a$ is an integration constant.
With thie (\ref{scfoeq}) is reduced to
\bea
&\dot \rho \pm \dfrac{m}{\varrho}(A
-\dfrac{\varrho^{2m}}{\varrho^{2m} + a^2}) \rho =0,  \nn\\
&\dfrac{m}{g} \dfrac{\dot A}{\varrho} \pm \dfrac{g}{2}(\rho^2-\rho_0^2) =0.
\eea
In this case the Hamiltonian becomes
\bea
{\cal H} = \dfrac{g}{2} H \rho_0^2,
\eea
and the energy (per unit length) acquires
the absolute minimum value
\bea
E = 2\pi m \rho_0^2 = \dfrac{g}{2} \rho_0^2 \hat \phi.
\eea
This tells that the minimum energy is fixed by the topological
flux quantum mumber.

\begin{figure}
\includegraphics[scale=0.7]{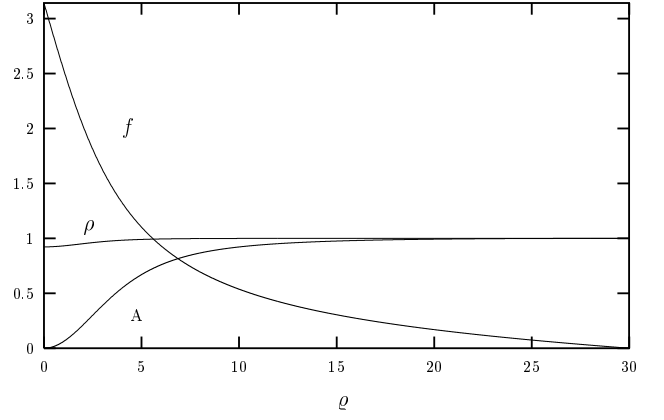}
\caption{The non-Abelian vortex with $m=1$ in two-gap superconductor
which has $2\pi/g$ flux. Notice that here we have $\rho(0)\ne 0$.}
\label{scv1}
\end{figure}

The magnetic vortex is topological. This is because 
the non-linear sigma field $\hn$ defined in (\ref{def})
naturally describes the mapping $\pi_2(S^2)$ from the compactified 
$xy$-plane $S^2$ to the physical target space $S^2$,
whose quantum number $\pi_2(S^2)$ is given by
\bea
&q = - \dfrac {1}{4\pi} \int \epsilon_{ij} \partial_i \zeta^{\dagger}
\partial_j \zeta d^2 x  \nn\\
&= \dfrac {1}{8\pi} \int \epsilon_{ij} \hn \cdot (\partial_i \hn
\times \partial_j \hn) d^2 x 
= \dfrac {m}{2} \int \dot f \sin f d \varrho \nn\\
&= m.
\label{scqn}
\eea
Clearly this topology is due to the non-Abelian nature 
of two-gap superconductor.
As we will see this is the topological quantum number 
that we will encounter repeatedly in the following.

We have shown that the above magnetic vortex has a non-Abelian flux
quantization rule. So one might wonder whether the two-gap
superconductor does not allow a magnetic vortex which
satisfies the ordinary flux quantization rule. 
It does. Indeed with a different boundary condition
\bea
&\dot \rho (0) = 0,~~~\rho(\infty) = \rho_0,
~~~f (0) = \pi,~~~f (\infty) = 0, \nn\\
& A (0) = 0,~~~A (\infty) = 1,
\label{scbc1}
\eea
we obtain another magnetic vortex solution in two-gap superconductor
which has $2\pi/g$ flux. This is shown in
Fig.\ref{scv1}. The difference between 
the two solutions is that in Fig.\ref{scv1}
both components of $\phi$ vanish at the core of the vortex, 
but in Fig.\ref{scv1} the second component has a finite
condensation due to the boundary condition
$\dot \rho(0)=0$. Notice that this boundary condition 
would have been unacceptable in ordinary
(one-gap) superconductor because it creates a physical singularity 
at the vortex core. But in two-gap superconductor this 
boundary condition produces a perfectly regular vortex. 
This is another unexpected feature of two-gap superconductor.

So far we have assumed the potential (\ref{scpot})
to obtain the vortex solution for simplicity. 
But as we have remarked the global $SU(2)$ symmetry 
of the potential will often be broken in reality \cite{maz}. 
For example the potential could be such that the vacuum density 
of two condensates is completely independent. In this case 
we may not be able to impose the boundary condition (\ref{scbc}),
in particular $f(\infty)=0$, and will no longer have 
the non-Abelian flux quantization \cite{maz,baba}.
This, together with the above result, tells that 
in two-gap superconductor different
boundary conditions lead to different flux quantizations.

\section{Comparison with Abelian Vortex}

At this point it is important to understand exactly
how different is the non-Abelian magnetic vortex from
the well-known Abelian Abrikosov
vortex \cite{abri}. 
To compare the non-Abelian vortex with
the Abelian vortex, we let $\phi$ be the charged scalar field 
of the electron-pair condensate in ordinary superconductor.
Then the Abrikosov vortex is described by 
the Abelian Landau-Ginzburg Lagrangian
\bea
&{\cal L} = - |D_\mu \phi|^2 + \mu^2 \phi^{\dagger}\phi
- \dfrac{\lambda}{2} (\phi^{\dagger} \phi)^2 
- \dfrac{1}{4} F_{\mu \nu}^2, \nn\\
&D_\mu \phi = (\partial_\mu + ig A_\mu) \phi,
\label{alag}
\eea
which has the equation of motion
\bea
&D^2\phi =\lambda(\phi^{\dagger} \phi
-\dfrac{\mu^2}{\lambda})\phi, \nn\\
&\partial_\mu F_{\mu \nu} = -j_\nu =  i g \Big[(D_\nu
\phi)^{\dagger}\phi - \phi ^{\dagger}(D_\nu \phi) \Big].
\label{aeq1}
\eea
With the ansatz
\bea
&\phi =\dfrac{1}{\sqrt 2} \rho(\varrho) \exp (-im\varphi), \nn\\
&A_\mu = \dfrac{m}{g} A(\varrho) \partial_\mu \varphi.
\label{aans}
\eea
(\ref{aeq1}) is reduced to
\bea
&\ddot{\rho}+\dfrac{1}{\varrho}\dot\rho - \dfrac{m^2}{\varrho^2}(A-1)^2 \rho
= \dfrac{\lambda}{2}(\rho^2-\rho_0^2)\rho, \nn\\
&\ddot A - \dfrac{1}{\varrho} \dot A - g^2 \rho^2 (A-1) =0.
\label{aeq2}
\eea
Now, with the boundary condition
\bea
&\rho (0) = 0,~~~\rho(\infty) = \rho_0 , \nn\\
&A (0) = 0, ~~~~~A (\infty) = 1,
\label{abc}
\eea
one can easily obtain the well-known Abelian Abrikosov
vortex solution \cite{abri}, whose magnetic flux is given by
\bea
&\hat \phi = \dfrac{}{} \int H d^2x
= \dfrac{2\pi m}{g} \big[A (\infty) - A (0) \big] \nn\\
&= \dfrac{2\pi m}{g}.
\label{aflux}
\eea
The ansatz (\ref{aans}) assures that the topological flux
quantum number of the Abelian vortex is fixed by $\pi_1(S^1)$.
Remember that here the penetration length of the magnetic
field $\lambda_H$ and the coherence length of the condensate $\xi_C$
are also given by (\ref{coleng}).

To compare the above Abelian vortex with the non-Abelian one,
let us consider the critical case $\sqrt \lambda=g$.
In this case (\ref{aeq2}) is reduced to
the first order equations
\bea
&\dot \rho \pm \dfrac{m}{\varrho} (A-1) \rho = 0, \nn\\
&\dfrac{m}{g} \dfrac{\dot A}{\varrho}
\pm \dfrac{g}{2} (\rho^2 - \rho_0^2) = 0.
\label{aeq3}
\eea
This has to be compared with (\ref{scfoeq}). When $f=0$ (or $f=\pi$),
the two sets of equations become identical to each other.
This is not surprising because when $f=0$ (or $f=\pi$) one component of the
doublet becomes zero, which effectively reduces the doublet to a singlet.
This tells two things. First, the two-gap superconductor can also
(as it should) allow the Abelian vortex (with $f=0$ or $f=\pi$),
whose topology is fixed by $\pi_1(S^1)$. Notice, however, in this case
the non-Abelian topological quantum number (as well as the non-Abelian
magnetic flux) defined by (\ref{scqn}) becomes identically zero.
Secondly, when $f(\varrho)$ has a non-trivial profile, there
is no way the non-Abelian vortex can be related to the Abelian one.
This is because in this case there is no way (no gauge transformation)
in which the doublet can be put into a singlet.
Moreover, the non-trivial $f(\varrho)$ ensures that the topology
of the non-Abelian vortex is $\pi_2(S^2)$, not $\pi_1(S^1)$.
This distinguishes our non-Abelian vortex from the Abelian
Abrikosov vortex.

Another important difference comes from the magnetic flux and the energy.
In the Abelian case the Hamiltonian, with (\ref{ccon}), becomes
\bea
{\cal H} = \dfrac{g}{2} H \rho_0^2,
\eea
so that the energy of the vortex (per unit length) is given by
\bea
E = \dfrac{g}{2} \int H \rho_0^2 d^2x = m\pi \rho_0^2
= \dfrac{g}{2} \rho_0^2 \hat \phi.
\eea
This means that the non-Abelian vortex has twice as much
magnetic flux and energy.
This difference can be traced back to the difference of the boundary
conditions (\ref{scbc}) and (\ref{abc}).
In the Abelian case (\ref{abc}) tells that $A(\infty)-A(0)=1$,
but in the non-Abelian case (\ref{scbc}) tells that
$A(\infty)-A(0)=2$. Obviously this difference makes the difference in
the magnetic flux and the energy. Mathematically this difference
originates from the fact that the Abelian $U(1)$ 
runs from $0$ to $2\pi$, but the $S^1$ fiber of $SU(2)$
runs from $0$ to $4\pi$.

\section{Non-Abelian Superconductor}

So far we have discussed an Abelian gauge theory of
two-gap superconductor. 
But notice that this type of superconductor must be made of
the doublet whose components carry the same charge
(e.g., a doublet made of two electron-electron pair
condensates or two hole-hole pair condensates),
because the doublet is coupled to the Abelian
electromagnetic field. Obviously the above theory can not
describe a doublet which is made of opposite charges 
(made of one electron-electron pair condensate and one hole-hole
pair condensate). In this case an Abelian gauge theory can not
explain the superconductivity. Now we show that
this type of two-gap superconductor can be explained by
a genuine non-Abelian $SU(2)$ gauge
theory, and show that this type of superconductor
also allows a non-Abelian magnetic vortex identical to 
what we have discussed before.

To construct a theory of superconductor which has
a genuine non-Abelian gauge symmetry, we need to understand
the mathematical structure of the non-Abelian gauge potential.
Consider $SU(2)$ and let $\hn$ be a gauge covariant unit triplet
which selects the charge direction of $SU(2)$. In this case
one can always decompose the
non-Abelian gauge potential into the restricted potential $\hat A_\mu$
and the valence potential $\X_\mu$ \cite{cho80,cho81},
\bea
& \vec{A}_\mu =A_\mu \n -
\oneg \n\times\pro_\mu\n+\X_\mu\nonumber
         = \hat A_\mu + \X_\mu, \nn\\
&  (A_\mu = \n\cdot \vec A_\mu,~ \n^2 =1,~
\hat{n}\cdot\vec{X}_\mu=0),
\label{cdecom}
\eea
where $ A_\mu$ is the
``electric'' potential. Notice that the restricted potential
is precisely the potential which leaves $\n$
invariant under parallel transport,
\bea
\D_\mu \n = \pro_\mu \n
+ g {\hat A}_\mu \times \n = 0. \eea Under the infinitesimal
gauge transformation \bea \delta \n = - \vec \alpha \times \n
\,,\,\,\,\, \delta \A_\mu = \oneg  D_\mu \vec \alpha, \eea one has
\bea &&\delta A_\mu = \oneg \n \cdot \pro_\mu \valpha,\,\,\,\
\delta \hat A_\mu = \oneg \D_\mu \valpha  ,  \nn \\
&&\hspace{1.2cm}\delta \X_\mu = - \valpha \times \X_\mu  .
\eea
This tells two things. First, $\hat A_\mu$ by itself describes an $SU(2)$
connection which enjoys the full $SU(2)$ gauge degrees of
freedom. Secondly, the valence potential $\vec X_\mu$ forms a
gauge covariant vector field under the gauge transformation.
Furthermore this tells that the decomposition is
gauge-independent. Once the gauge covariant topological field
$\hat n$ is given, the decomposition follows automatically
independent of the choice of a gauge \cite{cho80,cho81}.

The importance of the decomposition (\ref{cdecom}) for our
purpose is that one can construct a non-Abelian gauge theory,
a restricted gauge theory which has a full non-Abelian gauge
degrees of freedom, with the restricted potential
$\hat A_\mu$ alone \cite{cho80,cho81}. This is because
the valence potential $\vec X_\mu$ can be treated as
a gauge covariant source, so that one can exclude it from
the theory without compromizing the gauge invariance.
Indeed we will see that it is this
restricted gauge theory which describes the non-Abelian gauge
theory of superconductivity.

Remarkably the restricted potential $\hat{A}_\mu$ retains
all the essential topological characteristics of the original
non-Abelian potential.
First, $\hat{n}$ defines $\pi_2(S^2)$
which describes the non-Abelian monopoles \cite{cho80,cho81}.
Secondly, it characterizes $\pi_3(S^3)$ which describes
not only the topologically distinct vacua but also the instanton
numbers \cite{bpst,cho79}.
Furthermore it has a dual structure,
\begin{eqnarray}
& \hat{F}_{\mu\nu} = (F_{\mu\nu}+ H_{\mu\nu})\hat{n}\mbox{,}\nonumber \\
& F_{\mu\nu} = \partial_\mu A_{\nu}-\partial_{\nu}A_\mu \mbox{,}\nonumber \\
& H_{\mu\nu} = -\dfrac{1}{g} \hat{n}\cdot(\partial_\mu
\hat{n}\times\partial_\nu\hat{n}) = \partial_\mu
C_\nu-\partial_\nu C_\mu,
\end{eqnarray}
where $C_\mu$ is the ``magnetic'' potential \cite{cho80,cho81}.

With these preliminaries we now establish a
non-Abelian superconductivity. Consider a $SU(2)$ gauge theory
described by the Lagrangian in which a doublet $\Phi$ couples
to the restricted $SU(2)$ gauge potential,
\bea
&{\cal L} = -|\hat D_\mu \Phi|^2 + \mu^2 \Phi^{\dagger}\Phi
- \dfrac{\lambda}{2} (\Phi^{\dagger} \Phi)^2
-\dfrac{1}{4} {\hat F}_{\mu\nu}^2, \nn\\
&\hat D_\mu \Phi = ( \partial_\mu + \dfrac{g}{2i} \vec \sigma
\cdot \hat A_\mu ) \Phi. 
\label{nasclag1}
\eea
The equation of
motion of the Lagrangian is given by
\bea
&{\hat D}^2\Phi
=\lambda(\Phi^{\dagger} \Phi
-\dfrac{\mu^2}{\lambda})\Phi, \nn\\
&\hat D_\mu \hat F_{\mu \nu} = \vec j_\nu
= g \Big[(\hat D_\nu \Phi)^{\dagger}\dfrac{\vec\sigma}{2i} \Phi
- \Phi ^{\dagger} \dfrac{ \vec\sigma}{2i}(\hat D_\nu \Phi) \Big].
\label{nasceq}
\eea
Let $\xi$ and $\eta$
be two orthonormal doublets which form a basis,
\bea
&\xi{\dagger}\xi =1,~~~~~\eta^{\dagger} \eta = 1,
~~~~~\xi^{\dagger}\eta = \eta^{\dagger} \xi = 0, \nn\\
&\xi^{\dagger} \vec \sigma\xi = \hat n,
~~~~~\eta^{\dagger} \vec \sigma\eta= -\hat n, \nn\\
&(\hn \cdot \vec \sigma) ~\xi = \xi,
~~~~~(\hn \cdot \vec \sigma) ~\eta = -\eta,
\label{odbasis}
\eea
and let
\bea
\Phi = \phi_+ \xi + \phi_- \eta, ~~~~~(\phi_+= \xi^{\dagger}\Phi,
~~~\phi_-= \eta^{\dagger} \Phi).
\label{phidecom}
\eea
With this we have the identity
\bea
&\Big[\partial_\mu - \dfrac{g}{2i} \big(C_\mu \hn
+ \dfrac{1}{g} \hn \times \partial_\mu \hn \big)
\cdot \vec \sigma \Big] \xi = 0, \nn\\
&\Big[\partial_\mu + \dfrac{g}{2i} \big(C_\mu \hn
- \dfrac{1}{g} \hn \times \partial_\mu \hn \big)
\cdot \vec \sigma \Big] \eta = 0,
\label{cid0}
\eea
and find
\bea
\hat D_\mu \Phi = (D_\mu \phi_+) \xi + (D_\mu \phi_-)  \eta,
\label{phidef}
\eea
where
\bea
&D_\mu \phi_+ = (\partial_\mu + \dfrac{g}{2i} {\cal A}_\mu) \phi_+,
~~~D_\mu \phi_- = (\partial_\mu - \dfrac{g}{2i} {\cal A}_\mu) \phi_-, \nn\\
&{\cal A}_\mu = A_\mu + C_\mu, \nn\\
&C_\mu = \dfrac{2i}{g} \xi^{\dagger}\partial_\mu \xi
= - \dfrac{2i}{g} \eta^{\dagger}\partial_\mu \eta. \nn
\eea
From this we can express (\ref{nasclag1}) as
\bea
&{\cal L} = - |D_\mu \phi_+|^2 - |D_\mu \phi_-|^2
+ m^2 (\phi_+^{\dagger}\phi_+ + \phi_-^{\dagger}\phi_-) \nn\\
&- \dfrac{\lambda}{2} (\phi_+^{\dagger}\phi_+ + \phi_-^{\dagger}\phi_-)^2
-\dfrac{1}{4} {\cal F}_{\mu\nu}^2,
\label{nasclag2}
\eea
where
\bea
{\cal F}_{\mu\nu} = \partial_\mu {\cal A}_\nu - \partial_\nu {\cal A}_\mu, \nn
\eea
This tells that the restricted $SU(2)$ gauge theory
(\ref{nasclag1}) is reduced to
an Abelian gauge theory coupled to oppositely charged
scalar fields $\phi_+$ and $\phi_-$. We emphasize that
this Abelianization is achieved without any gauge fixing.

The Abelianization assures that the non-Abelian theory
is not different from the two-gap Abelian theory.
Indeed with
\begin{equation}
\chi = \left(\begin{array}{rr}
\phi_+\\
\phi_-^*
\end{array}\right),
\end{equation}
we can express the Lagrangian (\ref{nasclag2}) as
\bea
&{\cal L} = - |D_\mu \chi|^2 + \mu^2 \chi^{\dagger}\chi
- \dfrac{\lambda}{2} (\chi^{\dagger} \chi)^2
- \dfrac{1}{4} {\cal F}_{\mu \nu}^2, \nn\\
&D_\mu \chi = (\partial_\mu + ig {\cal A}_\mu) \chi,
\label{nalag}
\eea
This is formally identical to the Lagrangian (\ref{sclag}) of two-gap
Abelian superconductor discussed in Section II. The only difference is
that here $\phi$ and $A_\mu$ are replaced by $\chi$ and ${\cal A}_\mu$.
This establishes that, with the proper redefinition of field
variables (\ref{phidecom}) and (\ref{phidef}), our non-Abelian
restricted gauge theory (\ref{nasclag1}) can in fact be made
identical to the Abelian gauge theory of two-gap superconductor. This
proves the existence of non-Abelian superconductors
made of the doublet consisting of oppositely charged
condensates. As importantly our analysis tells that
the two-gap Abelian superconductor has a hidden
non-Abelian gauge symmetry because it can be transformed to
the non-Abelian restricted gauge theory. This implies that
the underlying dynamics of the topological superconductors
is indeed the non-Abelian gauge symmetry.
In non-Abelian superconductor it is explicit.
But in two-gap Abelian superconductor it is hidden,
where the full non-Abelian gauge symmetry only
becomes transparent when one embeds the nontrivial topology
properly into the non-Abelian symmetry \cite{cho2}.

Once the equivalence of two Lagrangians (\ref{sclag})
and (\ref{nasclag1}) is established, it must be evident
that the non-Abelian gauge theory of two-gap superconductor also
admits a non-Abelian magnetic vortex.
This proves the existence of a non-Abelian
Meissner effect and non-Abelian superconductivity. 
All the results of Section II become
equally valid here.

\section{Non-Abelian Vortex in Gross-Pitaevskii Theory 
of Two-component BEC}

The creation of the multi-component Bose-Einstein condensates
of atomic gases \cite{bec} has widely opened new opportunities for
us to study the topological objects experimentally which so far
have been only of theoretical interest. This is because
the multi-component BEC could have a complex
non-Abelian topological structure, and thus could have far more interesting
topological vortices. Already new vortices have
successfully been created with different methods in
two-component Bose-Einstein condensates \cite{exp1,exp2}.
But suprisingly, there have been few theoretical study of these
vortices. Indeed only recently the physical meaning of the vortices
has been clarified as a vorticity vortex \cite{bec5}.
In the following we discuss the non-Abelian vortex in
Gross-Pitaevskii theory of two-component BEC in detail.

Let a complex doublet $\phi=(\phi_1,\phi_2)$ be the two-component
BEC, and consider the non-relativistic
two-component Gross-Pitaevskii Lagrangian \cite{ruo,bat}
\bea
&{\cal L} = i \dfrac {\hbar}{2}
\Big[\big(\phi_1^\dag ( \partial_t \phi_1)
-( \partial_t \phi_1)^\dag \phi_1 \big) \nn\\
&+ \big(\phi_2^\dag ( \partial_t \phi_2)
-( \partial_t \phi_2)^\dag \phi_2 \big)  \Big]
- \dfrac {\hbar^2}{2M} (|\partial_i \phi_1|^2 + |\partial_i \phi_2|^2) \nn\\
& + \mu_1 \phi_1^* \phi_1 + \mu_2 \phi_2^* \phi_2
- \dfrac {\lambda_{11}}{2} (\phi_1^* \phi_1)^2 \nn\\
&- \lambda_{12} (\phi_1^* \phi_1)(\phi_2^* \phi_2)
- \dfrac {\lambda_{22}}{2} (\phi_2^* \phi_2)^2,
\label{gplag1}
\eea
where $\mu_i$ are the chemical potentials and 
$\lambda_{ij}$ are the quartic coupling constants which
are determined by the scattering lengths $a_{ij}$
\bea
\lambda_{ij}=\dfrac{4\pi {\hbar}^2}{M} a_{ij}.
\eea
Notice that here we have neglected the trapping
potential. This is justified if the range of the trapping
potential is much larger than the size of topological objects we
are interested in, and this is what we are assuming here. 

The Lagrangian has the global $U(1)\times U(1)$ symmetry. 
But notice that for the spin $1/2$ condensate of $^{87}{\rm Rb}$ atoms,
the scattering lengths $a_{ij}$ differ by only about
$3~\%$ or so \cite{exp1,exp2}. In this case one may safely assume
\bea
\lambda_{11} \simeq \lambda_{12} \simeq \lambda_{22} 
\simeq \bar \lambda.
\label{qint}
\eea
With this (\ref{gplag1}) can be written as
\bea
&{\cal L} = i\dfrac {\hbar}{2} \Big[\phi^\dag (\partial_t \phi)
-(\partial_t \phi)^\dag \phi \Big]
- \dfrac {\hbar^2}{2M} |\partial_i \phi|^2 \nn\\
&-\dfrac{\bar \lambda}{2} \big(\phi^\dag \phi -\dfrac{\mu}{\bar
\lambda} \big)^2 - \delta \mu \phi_2^* \phi_2,
\label{gplag2}
\eea
where
\bea
\mu=\mu_1,~~~~~\delta \mu = \mu_1-\mu_2.
\eea
Clearly the Lagrangian has a global $U(2)$ symmetry when
$\delta \mu=0$. So the $\delta \mu$ interaction can be understood
to be the symmetry breaking term which breaks the global $U(2)$
symmetry to $U(1)\times U(1)$. Physically $\delta \mu$ represents
the difference of the chemical potentials between $\phi_1$ and
$\phi_2$ (Here one can always assume $\delta \mu \geq 0$ without
loss of generality), so that it vanishes when the two condensates
have the same chemical potential. Even when they differ the
difference could be small, in which case the symmetry breaking
interaction could be treated perturbatively. This tells that the
theory has an approximate global $U(2)$ symmetry, even in the
presence of the symmetry breaking term \cite{bec5}. 
This confirms that the theory of two-component BEC
is essentially non-Abelian.

Normalizing $\phi$ to $(\sqrt{2M}/\hbar)\phi$ and putting
\bea
\phi = \dfrac {1}{\sqrt 2} \rho \zeta , ~~~~~(\zeta^\dag \zeta = 1)
\label{phi}
\eea
we have the following Hamiltonian
in the static limit (in the natural unit $c=\hbar=1$) from 
the Lagrangian (\ref{gplag2}),
\bea
&{\cal H} =  \dfrac {1}{2} (\pro_i \rho)^2 +
\dfrac {1}{2} \rho^2 |\pro_i \zeta|^2
+ \dfrac{\lambda}{8} (\rho^2-\rho_0^2)^2 \nn\\
&+ \dfrac{\delta \mu^2}{2} \rho^2 \zeta_2^*\zeta_2,
\label{gpham1}
\eea
where
\bea
&\lambda=4M^2 \bar
\lambda,~~~~~\rho_0^2=\dfrac{4\mu M}{\lambda}, ~~~~~\delta
\mu^2=2M\delta \mu.  \nn
\eea
This can be expressed as
\bea
&{\cal H} = \lambda \rho_0^4 ~{\hat {\cal H}}, \nn\\
&{\hat {\cal H}} = \dfrac {1}{2} (\hat \pro_i \hat \rho)^2 +
\dfrac {1}{2} \hat \rho^2 |\hat \pro_i \zeta|^2
+ \dfrac{1}{8} (\hat \rho^2-1)^2 \nn\\
&+ \dfrac{\delta \mu}{4\mu} \hat \rho^2 \zeta_2^*\zeta_2,
\label{gpham2}
\eea
where
\bea
&\hat \rho = \dfrac {\rho}{\rho_0},
~~~~~\hat \pro_i =\kappa \pro_i,
~~~~~\kappa = \dfrac {1}{\sqrt \lambda \rho_0}. \nn
\eea
Notice that ${\hat {\cal H}}$ is completely dimensionless, with
only one dimensionless coupling constant $\delta \mu/\mu$. This
tells that the physical unit of the Hamiltonian is $\lambda
\rho_0^4$, and the physical scale of the coordinates is
$\kappa$. Since the correlation length $\bar \xi$ is given by
$\bar \xi= 1/\sqrt {2\mu M}$, $\kappa$ is comparable to 
the correlation length ($\bar \xi=\sqrt 2 ~\kappa$). 

Minimizing the Hamiltonian (\ref{gpham1}) we have
\bea
& \pro^2 \rho - |\pro_i \zeta|^2 \rho
=\Big (\dfrac{\lambda}{2} (\rho^2-\rho_0^2)
+ \delta \mu^2 (\zeta_2^* \zeta_2) \Big) \rho, \nn\\
&\Big\{(\pro^2 - \zeta^\dag \pro^2 \zeta) + 2 \dfrac {\pro_i
\rho}{\rho}(\pro_i - \zeta^\dag \pro_i\zeta) \nn\\
&+\delta \mu^2 (\zeta_2^* \zeta_2) \Big\} \zeta_1 = 0, \nn\\
&\Big\{(\pro^2 - \zeta^\dag \pro^2 \zeta) + 2 \dfrac {\pro_i
\rho}{\rho}(\pro_i - \zeta^\dag \pro_i\zeta) \nn\\
&-\delta \mu^2 (\zeta_1^* \zeta_1) \Big\} \zeta_2 = 0, \nn\\
&\zeta^\dag \pro_i(\rho^2\pro_i\zeta)
-\pro_i(\rho^2\pro_i\zeta^\dag) \zeta =0.
\label{gpeq1}
\eea
The equation is almost identical to
the equation (\ref{sceq3}) of two-gap superconductor,
although on the surface it appears totally different from
(\ref{sceq3}). To show this we let
\bea
&\hn=\zeta^{\dagger} \vec \sigma \zeta, \nn\\
&V_\mu= -i \zeta^{\dagger} \pro_\mu \zeta,
\label{hn}
\eea
and find the following identities
\bea
&(\pro_\mu \hn)^2 = 4 \Big( |\pro_\mu \zeta|^2
- |\zeta^\dag \pro_\mu \zeta|^2\Big)
=4 \big(|\pro_\mu \zeta|^2 - V_\mu^2 \big), \nn\\
&\hn \cdot (\pro_\mu \hn \times \pro_\nu \hn) = -2i (\pro_\mu
\zeta^\dag \pro_\nu \zeta
- \pro_\nu \zeta^\dag \pro_\mu \zeta) \nn\\
&=2(\pro_\mu V_\nu - \pro_\nu V_\mu).
\label{nid}
\eea
Moreover, from (\ref{cid}) we have 
\bea
&\big(\pro_\mu -i V_\mu -\dfrac{1}{2} \vec \sigma 
\cdot \pro_\mu \hn \big) \zeta =0.
\label{cid1}
\eea
With these identities we can rewrite the equation (\ref{gpeq1}) 
into a completely different form.
Indeed with (\ref{nid}) the first equation of
(\ref{gpeq1}) can be written as
\bea
& \pro^2 \rho - \big[\dfrac{1}{4} (\pro_i \hn)^2 + V_i^2 \big] \rho
=\big[(\dfrac{\lambda}{2} (\rho^2-\rho_0^2) \nn\\
&+ \delta \mu^2 (\zeta_2^* \zeta_2) \big] \rho.
\eea
Moreover, with (\ref{cid1}) the second and third equations of
(\ref{gpeq1}) can be expressed as
\bea
&\dfrac{1}{2} \Big(A+\vec B \cdot \vec \sigma \Big) \zeta =0,  \nn\\
&A= \pro_\mu \hn^2+i (2 \zeta_2^* \zeta_2 - 1) \delta \mu^2, \nn\\
&\vec{B}= \pro^2 \hn + 2\dfrac{\pro_i \rho}{\rho} \pro_i \hn
+2i V_i \hn \times \pro_i \hn \nn\\
&+i \delta \mu ^2 \hat k,
\label{gpeq2b1}
\eea
where $\hat k=(0,0,1)$. 
Thus we can write (\ref{gpeq2b1}) as
\bea
&\vec{n}\times \partial ^2\vec{n}+2\dfrac{\partial _i\rho }\rho
\vec{n}\times \partial_i\vec{n}-2V_i\partial _i\vec{n} \nonumber \\             &=\delta \mu ^2 \hat k \times \hn.
\label{gpeq2b2}
\eea
Finally, the last equation of (\ref{gpeq1}) is written as
\bea
\partial_i (\rho^2 V_i) = 0,
\eea
which tells that $\rho^2 V_i$ is
solenoidal (i.e., divergenceless). So we can always
replace $V_i$ with another field $B_i$
\bea
&V_i= \dfrac {1}{\rho^2} \epsilon_{ijk} \pro_j B_k
=\dfrac {1}{\rho^2} \pro_j G_{ij}, \nn\\
&G_{ij}=\epsilon_{ijk} B_k,
\eea
and express (\ref{gpeq2b2}) as
\bea
&\hn \times \pro^2 \hn + 2\dfrac{\pro_i \rho}{\rho}
\hn \times \pro_i \hn + \dfrac{2}{\rho^2}
\pro_i G_{ij} \pro_j \hn \nn\\
&=\delta \mu ^2\hat k \times \vec{n}.
\eea
With this (\ref{gpeq1}) can now be written as
\bea
& \pro^2 \rho - \big[\dfrac{1}{4}(\pro_i \hn)^2 + V_i^2 \big] \rho
=\big[\dfrac{\lambda}{2} (\rho^2-\rho_0^2) \nn\\
&+ \delta \mu^2 (\zeta_2^* \zeta_2) \big] \rho, \nn\\
&\hn \times \pro^2 \hn + 2\dfrac{\pro_i \rho}{\rho}
\hn \times \pro_i \hn + \dfrac{2}{\rho^2} \pro_i G_{ij} \pro_j \hn
=\delta \mu ^2\hat k \times \vec{n}, \nn\\
&\pro_i G_{ij}= - \rho^2 V_j.
\label{gpeq2}
\eea
This tells that (\ref{gpeq1}) can be transformed to
a completely different form which has a clear physical meaning.
The last equation tells that the theory has a conserved $U(1)$
current $j_\mu$,
\bea
j_\mu=\rho^2 V_\mu,
\eea
which is nothing but the Noether current
of the global $U(1)$ symmetry of the Lagrangian (\ref{gplag2}).
The second equation tells that the theory has another
partially conserved $SU(2)$ Noether current $\vec j_\mu$,
\bea
\vec j_\mu= \rho^2 \hn \times \pro_\mu \hn
- 2 \rho^2 V_\mu \hn,
\eea
which comes from the approximate $SU(2)$ symmetry
of the theory broken by the $\delta \mu$ term.
It also tells that there is one more $U(1)$ current
\bea
k_\mu= \hat k \cdot \vec j_\mu,
\eea
which is conserved even when $\delta \mu$ is not zero.
This is because the $SU(2)$ symmetry is broken down to
$U(1)$ when $\delta \mu$ is not zero.

More importantly this reveals that the Gross-Pitaevskii theory of
two-component BEC is closely related to the Landau-Ginzburg theory
of two-gap superconductor. Indeed, the equation 
(\ref{sceq3}) of two-gap superconductor and the equation
(\ref{gpeq2}) of two-component BEC acquire formally an identical form
when $\delta \mu=0$, except that here $gB_i$ and $F_{ij}$ is
replaced by $V_i$ and $-G_{ij}$.
This is really remarkable, but actually is not surprising.
This is because, when the electromagnetic interaction is 
switched off, the Landau-Ginzburg Lagrangian (\ref{sclag}) 
reduces to the Gross-Pitaevskii Lagrangian (\ref{gplag2})
in the limit $\delta \mu=0$ (in non-relativistic limit). 
In this case the two theories really become identical.

\begin{figure}
\includegraphics[scale=0.7]{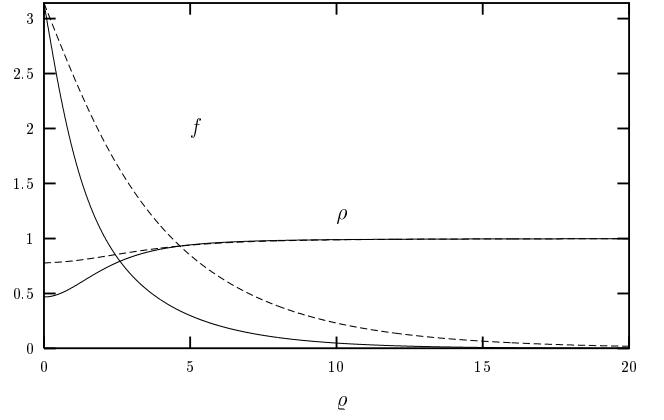}
\caption{The non-Abelian vortex in the Gross-Pitaevskii theory of
two-component BEC. Here we have put $n=1$, and $\varrho$ is in the
unit of $\kappa$. Dashed and solid lines correspond to $\delta
\mu/\mu= 0.1$ and $0.2$ respectively.}
\label{twobec-fig4}
\end{figure}

To obtain the vortex solution in two-component BEC, 
we choose the ansatz
\bea
&\rho= \rho(\varrho),  \nn\\
&\zeta = \Bigg( \matrix{\cos \dfrac{f(\varrho)}{2}
\exp (-im\varphi) \cr \sin \dfrac{f(\varrho)}{2} } \Bigg).
\label{gpans}
\eea
With the ansatz (\ref{gpeq1}) is reduced to
\bea
&\ddot{\rho}+\dfrac{1}{\varrho}\dot{\rho}-\bigg( \dfrac{1}{4}\dot{f}^2 
+\dfrac{m^2}{\varrho^2}\cos^2{\dfrac{f}{2}}
+\delta \mu^2 \sin^2{\dfrac{f}{2}} \bigg)\rho  \nn \\ 
&=\dfrac{\lambda}{2}(\rho^2-{\rho_0}^2)\rho,\nn \\
&\ddot{f}+ \bigg(\dfrac{1}{\varrho}
+2\dfrac{\dot{\rho}}{\rho}\bigg)\dot{f}
+\bigg( \dfrac{m^2}{\varrho^2}-\delta \mu^2\bigg)\sin{f} \nn \\
&=0.
\label{gpeq3}
\eea
Now with the boundary condition
\bea
&\dot \rho(0)=0,~~~\rho(\infty)=\rho_0,  \nn\\
&f(0)=\pi,~~~~~~f(\infty)=0,
\label{gpbc}
\eea
we can solve the equation and obtain the non-Abelian vortex 
solution in two-component BEC shown in Fig. \ref{twobec-fig4}. 
Notice that (\ref{gpeq3}) also admits
the well-known Abelian vortices with $\zeta_1=0$ 
or $\zeta_2=0$ (or equivalently $f=0$ or $f=\pi)$.
But obviously they are different from
the non-Abelian vortex, which has a non-trivial profile of 
$f(\varrho)$.

One can show that this vortex is topological, which carries 
a topological quantum number. In fact it can be viewed as 
a quantized vorticity flux \cite{cho1,bec5}. 
To see this notice that the potential $V_\mu$ defined
in (\ref{hn}) is nothing but the velocity field of
the doublet $\zeta$, which is given by
\bea
&V_\mu = -i\zeta^{\dagger} \pro_\mu \zeta
=-\dfrac{m}{2}(\cos{f}+1) \pro_\mu \varphi.
\label{gpvel}
\eea
This generates the vorticity
\bea
&\bar H_{\mu\nu}= \pro_\mu V_\nu - \pro_\nu V_\mu
=-i(\pro_\mu \zeta^{\dagger} \pro_\nu \zeta
-\pro_\nu \zeta^{\dagger} \pro_\mu \zeta) \nn\\
&=\dfrac{1}{2} \hn \cdot(\pro_\mu \hn \times \pro_\nu \hn) \nn\\ 
&=\dfrac{m}{2} \dot{f}\sin{f}(\pro_\mu \varrho \pro_\nu \varphi
-\pro_\nu \varrho \pro_\mu \varphi),
\label{gpvor}
\eea
which has a quantized vorticity flux
$\Phi_{\hat z}$ along the $z$-axis
\bea
&\Phi_{\hat z}=\dfrac{}{}
\int \bar H_{{\hat \varrho}{\hat \varphi}} \varrho d \varrho d \varphi
= -2\pi m.
\label{gpfluxz}
\eea
Furthermore, this flux can be viewed
to originate from a supercurrent which confines it
with a built-in Meissner effect
\bea
&\bar j_\mu = \pro_\mu \bar H_{\mu\nu} \nn\\
&=-m \big(\ddot f + \dfrac{\cos f}{\sin f} \dot f^2 
- \dfrac{1}{\varrho} \dot f \big) \sin f \partial_{\mu}\varphi, \nn\\
&\pro_\mu \bar j_\mu = 0.
\label{gpsc}
\eea
This tells that the vortex is a quantized vorticity flux
which is confined by the above supercurrent \cite{bec5}. 

To understand the topological nature of the vortex, notice that 
the vorticity $\bar H_{\mu\nu}$ is completely fixed by
the non-linear sigma field $\hn$, whose target space forms $S^2$. 
So $\hn$ naturally defines a mapping
$\pi_2(S^2)$ from the compactified $xy$-plane $S^2$ to
the target space $S^2$. The topological
quantum number of this mapping is given by
\bea
&q =  - \dfrac {i}{4\pi} \int \epsilon_{ij} \bar H_{ij} d^2 x \nn\\
&= - \dfrac {i}{4\pi} \int \epsilon_{ij} \partial_i \zeta^{\dagger}
\partial_j \zeta  d^2 x = m.
\label{gpqn}
\eea
This, of course, is mathematically identical to
the quantum number (\ref{scqn}) of 
the magnetic flux in two-gap superconductor. 
This confirms that the magnetic vortex in two-gap superconductor
and the vorticity vortex in two-component BEC have identical
topology. The only difference is that here the vortex 
describes a vorticity vortex, not a magnetic vortex.

There are three points which should be emphasized here.
First, the boundary condition (\ref{gpbc})
assures that at the vortex core we have only $\phi_2$
but at the infinity $\phi_1$ takes over completely.
This confirms that the vortex is essentially
non-Abelian. Secondly, when $\delta \mu$ approaches zero,
the vortex size becomes infinite so that it changes the shape 
of the vortex. But as far as it remains finite,
it does not change the physical nature of the vortex. 
In particular, the vortex can be interpreted as a vorticity flux. 
Finally, the vortex is topological in origin. 
It has a well-defined non-Abelian topology $\pi_2(S^2)$, 
even when $\delta \mu$ is not zero \cite{bec5}.
This is unexpected, because in this case we have only
$U(1)\times U(1)$ symmetry which can not provide
the non-Abelian topology.
Moreover the topological quantum number has a clear
physical meaning. It represents the vorticity quantum number.

In this paper we have adopted the $SU(2)$ symmetric
quartic interaction with (\ref{qint}) for simplicity. 
But in reality one may have to face a more complicated quartic 
interaction in two-component BEC. 
Nevertheless many of our 
results would undoubtedly survive with this complication.
For other quartic potentials
we refer the readers to existing literature \cite{bat,gar}.

\section{Gauge theory of Two-component BEC}

An important difference between two-component BEC and 
ordinary (one-component) BEC is the vorticity. 
The ordinary BEC has no vorticity but two-component BEC
has a non-trivial vorticity. This is because 
the velocity field of one-component BEC is given by the gradient
of the phase angle of the condensate, so that it  has
a vanishing vorticity. Due to the doublet structure, however,
the velocity field of two-component BEC is not given by the gradient
of the phase angle of the condensate. So a two-component BEC 
has a non-vanishing vorticity which plays an important role
as we have seen in the above analysis. 

In general creating vorticity costs energy.
But the above theory of two-component
BEC does not reflects this point properly, 
because the Gross-Pitaevskii Lagrangian (\ref{gplag1}) has no
vorticity interaction. But we can construct a gauge theory
of two-component BEC which can naturally accommodate the vorticity
interaction \cite{cho1}.
In the following we show that the gauge theory
of two-component BEC also allows 
a vorticity vortex very similar to the one we have in
Gross-Pitaevskii theory.

Consider a ``charged'' two-component condensate $\phi$ interacting
``electromagnetically'', which can be described by the
gauged Gross-Pitaevskii Lagrangian
\bea
&{\cal L} = i \dfrac {\hbar}{2} \Big[\phi^\dag (\tilde D_t \phi)
-(\tilde  D_t \phi)^\dag \phi \Big] 
- \dfrac {\hbar^2}{2M} |\tilde D_i \phi|^2 \nn\\
& + \mu \phi^\dag \phi - \dfrac {\lambda}{2}
(\phi^\dag \phi)^2 - \dfrac {1}{4} \tilde F_{\mu \nu} ^2, \nn\\
&\tilde D_\mu \phi = ( \pro_\mu - i g \tilde A_\mu ) \phi.
\label{beclag}
\eea
Notice that here we have assumed $\delta \mu=0$, so that
the global $U(2)$ symmetry of
the Lagrangian (\ref{gplag2}) is modified to a local $U(1)$
and a global $SU(2)$ symmetry.
Normalizing $\phi$ to $(\sqrt {2M}/\hbar) \phi$ 
and putting 
\bea
\phi=\dfrac {1}{\sqrt 2} \rho \zeta,
~~~~~(\zeta^{\dagger}\zeta=1) \nn
\eea
we have the following Hamiltonian from
(\ref{beclag}) in the static limit
\bea
&{\cal H} =  \dfrac {1}{2} (\pro_i \rho)^2
+ \dfrac {1}{2} \rho^2 |\tilde D_i \zeta |^2
+ \dfrac{\lambda}{8} (\rho^2 - \rho_0^2)^2,  \nn\\
&+ \dfrac {1}{4} \tilde F_{ij} ^2,
\label{becham1}
\eea
where $\rho_0^2=2\mu/\lambda$, and we have again
rescaled $\mu^2$ and $\lambda$.

Obviously the Lagrangian (\ref{beclag}) can be identified as
a non-relativistic Landau-Ginzburg Lagrangian of
two-gap superconductor. But, of course, 
here we are dealing with the neutral condensates,
so that the ``electromagnetic''
interaction should be treated not as independent but as
self-induced. This means that the gauge
potential has to be a composite field of the condensate,
and we may identify the ``electromagnetic'' potential
by the velocity field of $\zeta$ \cite{cho1},
\bea
&g \tilde A_\mu = V_\mu = -i\zeta^\dag \pro_\mu \zeta.
\label{am}
\eea
A justification for this is that we can actually derive this
from the Hamiltonian (\ref{becham1}) if we
neglect the Maxwell term. Indeed (\ref{am}) becomes
nothing but the Euler-Lagrange equation of the Hamiltonian
for the potential in the absence of the Maxwell term.

Now, introducing a $CP^1$ field $\xi$ by 
\bea
\zeta=\exp(i\gamma) \xi,~~~\xi^{\dagger}\xi=1,
\label{cp1}
\eea
we have
\bea
&\tilde A_\mu =-\dfrac{i}{g} \xi^\dag \pro_\mu \xi + \pro_\mu \gamma, \nn\\
&\tilde F_{\mu\nu} = -\dfrac{i}{g} (\pro_\mu \xi^\dag \pro_\nu \xi
- \pro_\nu \xi^\dag \pro_\mu \xi ) \nn\\
&=\dfrac{1}{g} \bar H_{\mu\nu},
\label{vor}
\eea
where $\bar H_{\mu\nu}$ is the vorticity of the velocity potential
$V_\mu$. With this the Hamiltonian (\ref{becham1}) is written as
\bea
&{\cal H} = \dfrac {1}{2} (\pro_i \rho)^2 + \dfrac {1}{2} \rho^2
\Big(|\pro_i \xi |^2 - |\xi^\dag \pro_i \xi|^2 \Big)
+ \dfrac{\lambda}{8} (\rho^2 - \rho_0^2)^2 \nn\\
&+ \dfrac {1}{4 g^2} (\pro_i \xi^\dag \pro_j \xi
- \pro_j \xi^\dag \pro_i \xi)^2.
\label{becham2}
\eea
This tells two things. First the Hamiltonian naturally accommodates
the vorticity interaction. Secondly the doublet $\zeta$ is 
completely replaced by the $CP^1$ field $\xi$ in the Hamiltonian,
so that the theory becomes a theory of $CP^1$ field 
(coupled to a scalar field $\rho$). 

One might wonder why we need to include the vorticity interaction
in the Hamiltonian, when we do not have such interaction
in one-component BEC. The reason is that creating a vorticity
costs energy. So physically it makes sense
to keep the vorticity interaction in the Hamiltonian.
Moreover, here the coupling constant $g$ now
represents the strength of the vorticity interaction.
So we can always remove the vorticity interaction if
necessary, by putting $g=\infty$. This justifies
the vorticity interaction in the Hamiltonian \cite{cho1}.

Minimizing the Hamiltonian (with the constraint $\xi^\dag \xi=1$)
we have the following equation of motion
\bea
& \pro^2 \rho - \Big(|\pro_i \xi |^2 - |\xi^\dag \pro_i \xi|^2 \Big)
\rho = \dfrac{\lambda}{2} (\rho^2 - \rho^2_0) \rho,\nn \\
&\Big\{(\pro^2 - \xi^\dag \pro^2 \xi) + 2 (\dfrac {\pro_i
\rho}{\rho} - \xi^\dag \pro_i\xi)(\pro_i - \xi^\dag \pro_i \xi) \nn\\
&- \dfrac {2}{g^2 \rho^2} \Big(\pro_i (\pro_i
\xi^\dag \pro_j \xi - \pro_j \xi^\dag \pro_i \xi) \Big)
(\pro_j - \xi^\dag \pro_j \xi) \Big\} \xi \nn\\
&= 0.
\label{beceq1}
\eea
To understand the meaning of (\ref{beceq1}) notice that 
we can rewrite the identities (\ref{nid}) and (\ref{cid1}) as
\bea
& |\pro_\mu \xi|^2 - |\xi^\dag \pro_\mu \xi|^2
= \dfrac{1}{4} (\pro_\mu \hn)^2, \nn\\
&-i(\pro_\mu \xi^\dag \pro_\nu \xi - \pro_\nu \xi^\dag \pro_\mu \xi)
= \dfrac{1}{2} \hn \cdot (\pro_\mu \hn \times \pro_\nu \hn) \nn\\
&= \bar H_{\mu\nu},
\label{fmn}
\eea
and
\bea
&\big(\pro_\mu -ig \tilde A'_\mu -\dfrac{1}{2} \vec \sigma
\cdot \pro_\mu \hn \big) \xi =0, \nn\\
&\tilde A'_\mu= -\dfrac{i}{g} \xi^{\dag} \pro_\mu \xi.
\label{cid2}
\eea
With this (following the same procedure we adopted in 
the above section) we can rewrite (\ref{beceq1}) as
\bea
&\pro^2 \rho - \dfrac{1}{4} (\pro_i \n)^2 \rho
= \dfrac{\lambda}{2} (\rho^2 - \rho_0^2)
\rho, \nn \\
&\n \times \pro^2 \n + 2 \dfrac{\pro_i \rho}{\rho} \n \times \pro_i
\n + \dfrac{1}{g^2 \rho^2} \pro_i \bar H_{ij} \pro_j \n = 0.
\label{beceq2}
\eea
This is the equation of two-component BEC that
we are looking for. Notice that the second equation
assures that the theory has a conserved $SU(2)$ current,
which is a direct consequence of the global $SU(2)$ symmetry
of the Lagrangian (\ref{beclag}).

The equation (\ref{beceq2}) should be compared with the equation
(\ref{sceq3}) of two-gap superconductor and the equation
(\ref{gpeq2}) of two-component BEC, which are very similar 
to each other. In particular the similarity between (\ref{sceq3}) 
and (\ref{beceq2}) is unmistakable. Indeed, when
$B_\mu=0$, the first two 
equations of (\ref{sceq3}) reduce exactly 
the equation (\ref{beceq2}) of two-component
BEC. This tells that the gauge theory of two-component BEC
is almost identical to the Landau-Ginzburg theory of
two-gap superconductor. The only difference is that
in BEC the electromagnetic interaction becomes induced, because
here the condensate is neutral \cite{cho1,cho2}.

With (\ref{fmn}) the Hamiltonian (\ref{becham2}) acquires 
a remarkable form
\bea
&{\cal H} = \dfrac{1}{2} (\pro_i \rho)^2
+ \dfrac{\rho^2}{8} (\pro_i \hn)^2
+\dfrac{\lambda}{8} (\rho^2 - \rho_0^2)^2 \nn\\
&+ \dfrac{1}{16 g^2} (\pro_i \hn \times \pro_j \hn)^2 \nn\\
&= \lambda \rho_0^4 \hat {\cal H},
\label{becham3}
\eea
where
\bea
&\hat {\cal H} = \dfrac{1}{2} (\hat \pro_i \hat \rho)^2
+ \dfrac{\hat \rho^2}{8} (\hat \pro_i \hn)^2
+\dfrac{1}{8} (\hat \rho^2 - 1)^2 \nn\\
&+ \dfrac{\lambda}{16 g^2}
(\hat \pro_i \hn \times \hat \pro_j \hn)^2, \nn\\
&\hat \rho = \dfrac {\rho}{\rho_0},
~~~~~\hat \pro_i =\kappa \pro_i,
~~~~~\kappa = \dfrac {1}{\sqrt \lambda \rho_0}. \nn
\eea
So the theory can be written completely in terms of the non-linear
sigma field $\hn$ (and the scalar field $\rho$).
The reason for this is the Abelian gauge invariance
of (\ref{beclag}), which remains intact with
the self-induced interaction introduced by (\ref{am}).
And, just as in two-gap superconductor,
this Abelian gauge invariance reduces
the physical target space of $\zeta$ to the gauge orbit space
$S^2 = S^3/S^1$ of the $CP^1$ field $\xi$, 
which is identical to the target space of the non-linear 
sigma field $\n$. This is why we could transform the equation of motion 
(\ref{beceq1}) completely into the equation
for $\n$ in (\ref{beceq2}). 
This means that we can describe the theory as a self interacting
$CP^1$ model, or equivalently a self interacting
non-linear sigma model (coupled to a scalar 
field $\rho$) \cite{cho1}.

\begin{figure}
\includegraphics[scale=0.5]{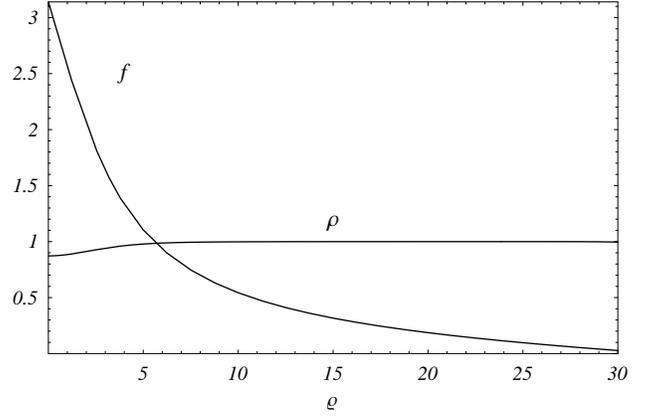}
\caption{The non-Abelian vortex with $m=1$ in gauge theory of
two-component BEC. Here we have put $g=\sqrt \lambda=1$, 
and $\varrho$ is in the unit of $\kappa$.}
    \label{becfig}
\end{figure}

To construct the desired vortex solution in this theory
we choose the ansatz
\bea
&\rho= \rho(\varrho), \nn\\
&\zeta= \Bigg( \matrix{\cos \dfrac{f(\varrho)}{2}
\exp (-im\varphi) \cr \sin \dfrac{f(\varrho)}{2} } \Bigg),  \nn\\
&\n= \Bigg(\matrix{\sin{f(\varrho)}\cos{m\varphi} \cr
\sin{f(\varrho)}\sin{m\varphi} \cr \cos{f(\varrho)}}\Bigg), \nn\\
&\tilde A_\mu = -\dfrac{m}{2g} \big(\cos{f(\varrho)}+1 \big) 
\partial_\mu \varphi,
\label{bvans}
\eea
and reduce the equation to
\bea
&\ddot{\rho}+\dfrac{1}{\varrho}\dot\rho
- \dfrac{1}{4}\Big(\dot{f}^2+\dfrac{m^2}{\varrho^2}\sin^2{f}\Big)\rho
= \dfrac{\lambda}{2}(\rho^2-\rho_0^2)\rho, \nn\\
&\Big(1+\dfrac{m^2}{g^2 \varrho^2}\dfrac{\sin^2{f}}{\rho^2}\Big) \ddot{f} \nn\\
&+ \Big( \dfrac{1}{\varrho}+ 2\dfrac{\dot{\rho}}{\rho}
+\dfrac{m^2}{g^2 \varrho^2} \dfrac{\sin{f}\cos{f}}{\rho^2} \dot{f}
- \dfrac{m^2}{g^2 \varrho^3}\dfrac{\sin^2{f}}{\rho^2} \Big) \dot{f} \nn\\
&- \dfrac{m^2}{\varrho^2}\sin{f}\cos{f}=0.
\label{bveq}
\eea
Notice the remarkable similarity between the above equation
and the magnetic vortex equation (\ref{sceq5}) in two-gap
superconductor. Indeed without the electromagnetic field
(i.e., with $B=0$) the first two equations of (\ref{sceq5})
becomes identical to the above equation.
Now, with the boundary condition
\bea
&\dot \rho(0)=0,~~~\rho(\infty)=\rho_0,  \nn\\
&f(0)=\pi,~~~~~~f(\infty)=0,
\label{becbc}
\eea
we obtain the the non-Abelian vortex solution in gauge theory
of two-component BEC shown in Fig.\ref{becfig}. 
Notice again that the non-trivial profile of $f$
assures that the vortex is non-Abelian.

Just like in the Gross-Pitaevskii theory the vortex here is 
topological, whose topological quantum number is expressed 
by $\pi_2(S^2)$ of the condensate $\xi$ (or by 
the non-linear sigma field $\hn$),
\bea
&q = - \dfrac {i}{4\pi} \int \epsilon_{ij} \partial_i \xi^{\dagger}
\partial_j \xi  d^2 x \nn\\
&= \dfrac {1}{8\pi} \int \epsilon_{ij} \hn \cdot
(\pro_i \hn \times \pro_j \hn) d^2 x = m.
\label{becqn}
\eea
Moreover, our analysis tells that the non-Abelian
vortex is nothing but the quantized vorticity flux $\bar H_{\mu\nu}$,
which is confined by a Meissner effect. Again this is because
the $U(1)$ gauge symmetry of (\ref{becham2})
assures the existence of a conserved supercurrent
\bea
\bar j_\mu=\partial_\nu \bar H_{\mu\nu},
~~~~~\partial_\mu \bar j_\mu=0,
\label{becsc}
\eea
which generates and confines the vorticity flux. 
This tells that two vortex solutions in this and
the last section are almost identical. Both
describe a quantized vorticity and have identical
topology. The only difference is the dynamics.
In the firse case the vorticity interaction is absent, but
in the second case it plays an important role.
So we have two competing theories of two-component
BEC. Which describes the real two-component BEC can only be
answered by experiments.

\section{A Prototype Non-Abelian Vortex: Baby Skyrmion}

There is a well-known prototype non-Abelian vortex, 
the baby skyrmion in Skyrme theory, which is very similar to 
the non-Abelian vortices in condensed matters.
In fact all non-Abelian vortices that we discussed in this paper
originate from the baby skyrmion. This is because Skyrme theory
itself is closely related to the above theories of
two-gap superconductor and two-component BEC.
In this sense it is worth reviewing the Skyrme theory
and clarify the connection between the Skyrme theory
and above theories of condensed matters.

The Skyrme theory has a rich topological
structure. The theory allows not only the original
skyrmion \cite{skyr}, but also a prototype knot 
known as Faddeev-Niemi knot \cite{cho01,fadd1}.
Furthermore, it allows
a non-Abelian vortex called the baby skyrmion \cite{piet}.
Since the Skyrme theory can be viewed as a non-linear
sigma model, it has a global $SU(2)$ symmetry which allows
a non-trivial topology $\pi_3(S^3)$. In (1+3)-dimension
this $\pi_3(S^3)$ is responsible for the skyrmion. 
But with the Hopf fibering of $S^3$ to $S^2 \times S^1$,
this $\pi_3(S^3)$ can be reduced
to $\pi_3(S^2)$ which provides the knot topology.
Moreover in (1+2)-dimension the theory allows $\pi_2(S^2)$,
which is responsible for the baby skyrmion.
Among these topological objects the baby skyrmion
and the knot are of particular relevance to us.

The importance of the baby skyrmion follows from the fact that
it can give rise to the Faddeev-Niemi knot.
In fact one can construct a twisted vortex ring by twisting it 
(making it periodic in $z$-coordinate) and connecting the periodic ends 
together. The twisted vortex ring acquires the knot
topology $\pi_3(S^2)$, and becomes 
the Faddeev-Niemi knot \cite{cho01,fadd1}.
This observation strongly implies the existence of topological knots in 
two-gap superconductor and two-component BEC, 
because one could also construct a twisted vortex ring with
the above non-Abelian vortices. 
So it is crucial to understand the baby skyrmion for us
to construct the non-Abelian knots in condensed matters.
For this reason we briefly review the baby skyrmion
in this section.

Let $\omega$ and $\hat n$ be the massless scalar field 
and non-linear sigma field in Skyrme theory. With
\bea
&U = \exp (\omega \dfrac{\vec \sigma}{2i} \cdot \hat n) \nn\\
&= \cos \dfrac{\omega}{2} - i (\vec \sigma \cdot \hat n)
\sin \dfrac{\omega}{2} ~~~~~({\hat n}^2 = 1), \nn\\
&L_\mu = U\partial_\mu U^{\dagger},
\label{su2}
\eea
one can write the Skyrme Lagrangian as
\bea
&{\cal L} = \dfrac{\mu^2}{4} {\rm tr} ~L_\mu^2 + \dfrac{\alpha}{32}
{\rm tr} \left( \left[ L_\mu, L_\nu \right] \right)^2.
\label{sklag}
\eea
A remarkable feature of the Skyrme Lagrangian is that
\bea
\omega = \pi,
\label{sfcon}
\eea
is a solution of the equation of motion, independent of $\hn$.
This means that, as far as we are concerned with the classical
solutions, we can assume $\omega = \pi$. In this case
the Lagrangian (\ref{sklag}) is reduced to the Skyrme-Faddeev
Lagrangian \cite{cho01},
\bea
{\cal L} \rightarrow -\dfrac{\mu^2}{2} (\partial_\mu
\hat n)^2-\dfrac{\alpha}{4}(\partial_\mu \hat n \times
\partial_\nu \hat n)^2,
\label{sflag}
\eea
whose equation of motion is given by
\bea
&\hn \times \partial^2 \hn + \dfrac{\alpha}{\mu^2} 
( \partial_\mu \tilde H_{\mu\nu} )
\partial_\nu \hn = 0, \nn\\
&\tilde H_{\mu\nu} = \hn \cdot (\partial_\mu \hn \times \partial_\nu \hn). 
\label{sfeq}
\eea
This is the equation which describes not only the baby skyrmion
but also the Faddeev-Niemi knot \cite{cho01}. 
Notice that here $\tilde H_{\mu\nu}$ is mathematically identical 
to the vorticity $\bar H_{\mu\nu}$ in BEC (up to the overall 
factor two), so that it can be expressed as a field strength of 
pootential $2V_\mu$. 

Now it must be clear that the equation (\ref{sfeq}) is very similar 
to the equation (\ref{sceq3}) of two-gap superconductor and 
the equations (\ref{gpeq2}) and (\ref{beceq2}) of two-component BEC. 
In fact all three equations of two-gap superconductor and
two-component BEC can be viewed as straightforward
generalizations of (\ref{sfeq}). This implies that 
the Skyrme theory and the above
theories of condensed matters are closely related.

\begin{figure}
\includegraphics[scale=0.7]{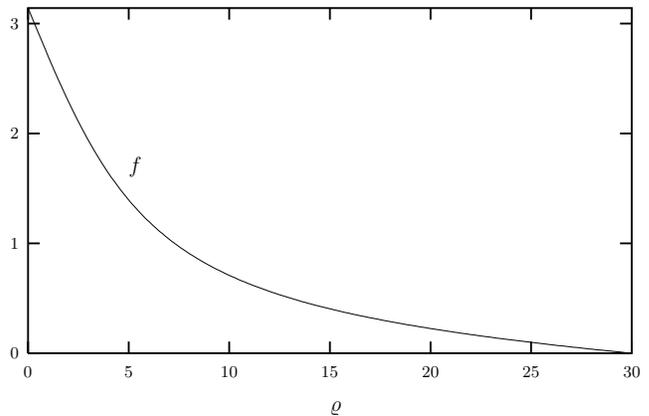}
\caption{The baby skyrmion with $m=1$ in Skyrme theory,
where we have put $\alpha/\mu^2=1$
for simplicity. Notice that $\varrho$ is in the unit of
$\sqrt \alpha/\mu$.}
\label{babysk}
\end{figure}

The Skyrme-Faddeev theory allows (not only the Faddeev-Niemi knot
but also) a vortex solution called the baby skyrmion \cite{piet}.
To be specific, let
\bea
\n=\Bigg(\matrix{\sin{f(\varrho)}\cos{m\varphi} \cr \sin{f(\varrho)}
\sin{m\varphi} \cr \cos{f(\varrho)}}\Bigg).
\label{bsans}
\eea
With this the equation (\ref{sfeq}) becomes
\bea
&\Big( 1+\dfrac{\alpha}{\mu^2}\dfrac{m^2}{\varrho^2}
\sin^2f \Big) \ddot{f} \nn\\
&+ \Big(\dfrac{1}{\varrho}
+\dfrac{\alpha}{\mu^2}\dfrac{m^2}{\varrho^2}
\dot{f} \sin{f}\cos{f} 
- \dfrac{\alpha}{\mu^2}\dfrac{m^2}{\varrho^3} \sin^2f \Big) \dot{f} \nn\\
&- \dfrac{m^2}{\varrho^2} \sin{f}\cos{f}=0.
\label{bseq}
\eea
Now, with the boundary condition
\bea
f (0) = \pi,~~~f (\infty) = 0,
\label{bsbc}
\eea
one can obtain a (massless) baby skyrmion solution
shown in Fig. \ref{babysk}. Notice 
the unmistakable similarity between the equation (\ref{bseq}) 
and the vortex equations (\ref{sceq5}) and (\ref{bveq}) of 
two-gap superconductor and two-component BEC. With 
$\rho=\rho_0$ the second equation of (\ref{bveq})
becomes identical to (\ref{bseq}). Furthermore with 
$\rho=\rho_0$ and $B=0$ the second equation of (\ref{sceq5})
becomes identical to (\ref{bseq}). In particular, when $\alpha=0$,
the above equation has the well-known
analytic solution
\bea
\cos f (\varrho) = \dfrac{\varrho^{2m} - a^2}{\varrho^{2m} + a^2},
~~~\sin f (\varrho) = \dfrac{2a\varrho^m}{\varrho^{2m} + a^2},
\eea
which is precisely the solution (\ref{bssol}) that we have in 
two-gap superconductor.

This implies that the baby skyrmion can actually
be viewed as a confined magnetic vortex in condensed matter,
confined by the Meissner effect. Indeed we can view
that $\tilde H_{\mu\nu}$ in (\ref{sfeq}) is generated by 
the conserved supercurrent
\bea
\tilde j_\mu=\partial_\nu \tilde H_{\mu\nu},
~~~~~\partial_\mu \tilde j_\mu=0.
\label{bssc}
\eea
This means that we can interprete $\tilde H_{\mu\nu}$ as a magnetic
field created by the supercurrent which confines 
the magnetic flux. This demonstrates the existence of
the Meissner effect in Skyrme theory. This tells 
that the baby skyrmion is very much like
a magnetic vortex in two-gap supercnductor or a vorticity
field in two-component BEC.

Clearly the baby skyrmion is topological. 
Again with the one-point compactification of the $xy$-plane $R^2$
to $S^2$, $\n$ defines the homotopy $\pi_2(S^2)$ of the mapping
from the $S^2$ to the target space $S^2=SU(2)/U(1)$ with the
integral topological quantum number $m$,
\bea
q=\dfrac{1}{8\pi} \int \epsilon_{ij} \hn \cdot
(\partial_i \hn \times \partial_j \hn) ~d^2x= m.
\label{bsqn}
\eea
Obviously this is identical to the topological quantum number
of the non-Abelian vortices in consensed matters. 
This confirms that indeed the baby skyrmion is 
a prototype non-Abelian vortex which is very similar to all
non-Abelian vortices in consensed matters. 

The baby skyrmion is unstable in the sense that, by
enlarging the size of the baby skyrmion, one can lower the energy
of the baby skyrmion. So it could not be thought
to represent a realistic physical object. Nevertheless it
plays a very important role because it can give 
rise to the Faddeev-Niemi knot \cite{cho01}.
Indeed one can view the knot as a twisted magnetic flux ring,
which one obtains by twisting the baby skyrmion (making it
periodic in $z$-coordinate) and connecting
the periodic ends together. The identification of the 
Faddeev-Niemi knot as a twisted vortex ring made of baby skyrmion
strongly indicates the existence of a similar knot in
two-gap superconductor and two-component BEC.
This is because we can also construct a similar knot 
by twisting the magnetic vortex or the vorticity vortex
to make a twisted vortex ring \cite{cho1,cho2,bec5}. 

The knot quantum number of Faddeev-Niemi knot
is given by 
\bea
&k=\dfrac{1}{16\pi^2} \int \epsilon_{ijk} V_i \tilde H_{jk} d^3x \nn\\
&=\dfrac{1}{4\pi^2} \int \epsilon_{ijk} \xi^{\dagger}\pro_i \xi
(\pro_j \xi^{\dagger} \pro_k \xi) d^3 x.
\eea
Exactly the same topology should describe the non-Abelian knots
in two-gap superconductor and two-component BEC.
 
\section{Discussions}

In this paper we have presented a convincing evidence how the new
condensed matters, in particular the two-gap
superconductor and two-component BEC, can have novel 
non-Abelian vortices. A characteristic feature of the non-Abelian vortices
is the non-trivial profile of the doublet. At the core the vortex is
made of only the second component, but as we move away
from the core the first component takes over and
fills the space completely at infinity.

Our anslysis tells that at
the center of all these non-Abelian vortices lies the baby skyrmion.
Indeed all these non-Abelian vortices stem from
the baby skyrmion. This suggests that the Skyrme theory could also
play an important role in condensed matter physics.
Ever since Skyrme proposed his theory, the Skyrme theory has always
been associated to nuclear and/or high energy physics.
This has lead people to believe that the topological objects
in Skyrme theory can only be realized at high energy,
at the $GeV$ scale. But our analysis opens up a new possibility
for us to construct them in a completely different environment
at much lower energy scale, in the new condensed
matters \cite{cho2,cho1}. This is really remarkable.

Another important lesson from our analysis is that the non-Abelian dynamics
could play a crucial role in condensed matter, in particular in
multi-component condensed matter. Perhaps this might not be
so surprising, given the fact that the multi-component condensed matters
can only (and naturally) be identified as non-Abelian multiplets.
Nevertheless, it is really remarkable that one can actually
construct a non-Abelian gauge theory of superconductivity.
Moreover the gauge theory of non-Abelian superconductivity 
suggests that, implicitly or explicitly, the underlying dynamics of 
multi-component condensed matters
can ultimately be related to a non-Abelian dynamics. This suggests that
the non-Abelian gauge theory could play an important 
role in condensed matter physics in the future.

Perhaps a most immediate outcome of our analysis
is the existence of topological knots in 
two-component condensed matters \cite{cho1,cho2}.
From our analysis it must become clear that
we can also construct a helical vortex
by twisting the non-Abelian vortex along the $z$-axis
and making it periodic in $z$-coordinate. Of course such a helical
vortex by itself may be unstable and likely to unwind itself to
a straight non-Abelian vortex, unless the periodicity
condition is enforced by hand. However, we can make it stable
by making it a vortex ring by smoothly connecting
two periodic ends. The stability follows from the fact that 
due to the twist the vortex ring can not collapse dynamically.
For example the twisted vorticity ring has a velocity
current along the knot. This in turn generates a non-vanishing 
angular momentum around the $z$-axis, which provides a 
centrifugal repulsive force to prevent the collapse of the 
vortex ring. So the twist which creates the instability in
the helical vortex now serves to provide the dynamical 
stability of the knot \cite{cho1,cho2}.

Furthermore, this dynamical stability of the knot can be
backed up by the topological stability. This is because
mathematically the non-linear sigma field $\hn$, after forming a knot,
acquires a non-trivial topology $\pi_3(S^2)$, 
which can not be changed by a smooth deformation of the field. 
This endorses our earlier claim that
the new condensates allow not only the non-Abelian vortices
but also stable knots \cite{cho2,cho1}.

From our analysis there should be no doubt that the non-Abelian
vortices and the topological knots must exist in the new
condensed matters. If so, the challenge now is
to verify the existence of these topological
objects experimentally. Constructing the knots
might not be a simple task at present moment. But the construction
of the non-Abelian vortices could be rather straightforward (at
least in principle), which might have already been done \cite{exp1,exp2}.
To identify the non-Abelian vortices, there are two points
one has to keep in mind. First, the (magnetic) flux of the
non-Abelian vortices is twice as much as that of the Abelian
counterparts. Secondly, the non-Abelian vortices must have
a non-trivial profile of $f(\varrho)$. This is a crucial
point which distinguishes them from the Abelian vortices.
With this in mind, one should be able to construct and identify
the non-Abelian vortices in the new condensates without much difficulty.

We conclude with the following remarks: \\
1. We have emphasized the potential importance of vorticity 
interaction in two-component BEC which is different from 
the polynomial Gross-Pitaevskii interaction which
one has in single-component BEC. The advantage of 
the vorticity interaction is that
it is a gauge interaction, except that here the gauge potential
is given by the velocity field of doublet $\zeta$.
This makes the gauge theory very similar to the 
Landau-Ginzburg theory of two-gap superconductor. 
In spite of the apparent similarity
the vortices in single-component BEC and in one-gap
superconductor have been thought to be based on seemingly different
dynamics, the one on the polynomial interaction the other on
the gauge interaction. Our self-interaction restores their
similarity at the theoretical level. We propose that the self-induced
gauge theory of two-component BEC could also play a fundamental
role in studying the non-Abelian superfluidity
in multi-component superfluid. In fact we believe that 
the vorticity vortex in two-component BEC could 
also describe the vortex in $\rm ^3He$ superfluid \cite{volo}.  \\
2. As we have pointed out, the two-component BEC and two-gap
superconductor can also admit the Abelian vortex
as a solution when $f=0$ or $f=\pi$.
Moreover, in the critical case when the coherence length and
the penetration length are the same, the magnetic flux and
the energy of the non-Abelian vortex with $q=m$ and those of
Abelian vortex with $q=2m$ becomes degenerate.
This raises an intriguing possibility, the possibility 
of tunneling between the Abelian vortex with $q=2m$
and the non-Abelian vortex with $q=m$. This is an interesting question
deserved to be studied further. \\
3. In this paper we have concentrated on
the condensed matter physics. But it must
be clear that our results should also have important implications
in high energy physics and cosmology. Indeed the existence of a 
chromoelectric knot in QCD \cite{plb05}
and a electroweak knot in Weinberg-Salam model \cite{ewknot}
which are very similar to the 
knots discussed here have already been proposed.
Moreover the existence of a cosmic string in cosmology 
has been speculated by many authors \cite{witt}.
We believe that the above theories of condensed matters, 
in particular the $SU(2)$ gauge theory of superconductor 
can easily be embedded
in any standard model of grand unification, and be
a reallistic model for such a cosmic string. 

{\bf ACKNOWLEDGEMENT}

~~~One of us (YMC) thanks G. Sterman for the kind hospitality during his
visit at C.N. Yang Institute for Theoretical Physics.
The work is supported in part by the ABRL Program of
Korea Science and Enginering Foundation (Grant R14-2003-012-01002-0) 
and by the BK21 Project of Ministry of Education.

\end{document}